\documentclass{tlp}

\usepackage{url,alltt,amsmath,epsfig,epsf,amssymb}
\usepackage[utf8]{inputenc}
\usepackage[all]{xy}
\usepackage{todo}

\newtheorem{theorem}{Theorem}
\newtheorem{corollary}[theorem]{Corollary}
\newtheorem{lemma}[theorem]{Lemma}

\newtheorem{example}{Example}[section]
\newtheorem{definition}{Definition}[section]

\newcommand{\st}[3]{\ensuremath{\langle #1 , #2 , #3 \rangle}}
\newcommand{\CT}{\ensuremath{\mathcal{CT}}}

\newcommand{\mcCP}{\ensuremath{\mathcal{CP}}}
\newcommand{\mcI}{\ensuremath{\mathcal{I}}}
\newcommand{\mcG}{\ensuremath{\mathcal{G}}}
\newcommand{\mcP}{\ensuremath{\mathcal{P}}}
\newcommand{\mcS}{\ensuremath{\mathcal{S}}}

\newcommand{\Ps}{\ensuremath{\mcP_1,\mcP_2}}
\newcommand{\Ss}{\ensuremath{\mcS_1,\mcS_2}}

\newcommand{\bbB}{\ensuremath{\mathbb{B}}}
\newcommand{\bbBca}{\ensuremath{\bbB_c \land \bbB_a}}
\newcommand{\bbG}{\ensuremath{\mathbb{G}}}
\newcommand{\bbV}{\ensuremath{\mathbb{V}}}

\newcommand{\subxt}{\left[x/t\right] }
\newcommand{\psubxt}{\left[x\wr t\right] }
\newcommand{\subxy}{\left[x/y\right] }

\newcommand{\extend}{\ensuremath{\lhd}}
\newcommand{\mcp}{\ensuremath{\mathcal{M}^\mcG_e(\sigma_{\mcCP})}}
\newcommand{\sigcp}{\ensuremath{\sigma_{\mcCP}}}

\newcommand{\der}{\ensuremath{\rightarrowtail}}
\newcommand{\derrev}{\ensuremath{\ {}^*\!\!\leftarrowtail}}
\newcommand{\dergts}{\ensuremath{\Rightarrow}}

\DeclareMathOperator{\var}{var}
\DeclareMathOperator{\vars}{vars}
\DeclareMathOperator{\dvar}{dvar}

\DeclareMathOperator{\type}{type}
\DeclareMathOperator{\src}{src}
\DeclareMathOperator{\tgt}{tgt}
\DeclareMathOperator{\gnd}{ground}
\DeclareMathOperator{\kp}{keep}
\DeclareMathOperator{\track}{tr}
\DeclareMathOperator{\id}{id}
\DeclareMathOperator{\node}{node}
\DeclareMathOperator{\edge}{edge}

\newcommand{\chr}{\ensuremath{{\tt chr}}}
\newcommand{\VARS}{\ensuremath{\mbox{VARS}}}
\newcommand{\chrrule}{\ensuremath{\varrho}}
\newcommand{\eqct}{\ensuremath{\stackrel{\CT}{\equiv}}}
\newcommand{\eqsubst}{\ensuremath{\stackrel{\text{Subst}}{\equiv}}}
\newcommand{\prodrule}{\ensuremath{p = (L \stackrel{l}{\leftarrow} K
\stackrel{r}{\rightarrow} R)\ }}
\newcommand{\strong}{\ensuremath{\mathcal{S}}}
\newcommand{\iso}{\ensuremath{\simeq}}

\submitted{25 September 2009}
\revised{18 April 2010}
\accepted{2 June 2010}

\begin{document}

\title{Analyzing Graph Transformation Systems through Constraint Handling Rules}
\shorttitle{Analyzing GTS through CHR}

\author[Frank Raiser, Thom Fr{\"u}hwirth]{FRANK RAISER \and THOM FR{\"U}HWIRTH\\
Faculty of Engineering and Computer Sciences, Ulm University, Germany\\
\email{\{Frank.Raiser|Thom.Fruehwirth\}@uni-ulm.de}}

\maketitle 


\begin{abstract}
Graph transformation systems (GTS) and constraint handling rules (CHR) are
non-deterministic rule-based state transition systems. 
CHR is well-known for its powerful confluence and program equivalence analyses,
for which we provide the basis in this work to apply them to GTS.
We give a sound and complete embedding of GTS in CHR, investigate confluence of
an embedded GTS, and provide a program equivalence analysis for GTS via the
embedding.
The results confirm the suitability of CHR-based program analyses for
other formalisms embedded in CHR.
\end{abstract}


\section{Introduction}
\label{sec:intro}

Graph transformation systems (GTS) are used to describe complex structures and
systems in a concise, readable, and easily understandable way. They have
applications ranging from implementations of programming languages over model
transformations to graph-based models of computation
\cite{Blostein1995,ehrigprangetaentzer06}. Graph transformation systems see
widespread use in many applications \cite{ehrigprangetaentzer06}, and hence
performing program analysis on them is becoming more important.

Constraint handling rules (CHR) \cite{fruehwirth09} on the other side allows for
rapid prototyping of constraint-based algorithms. Besides constraint reasoning,
CHR has been used for such diverse applications as type system design for Haskell
\cite{sulz_schr_stuck_aplas06}, time tabling
\cite{abd_marte_timetabling_chr_aai00}, computational linguistics
\cite{dahl_maharshak_dna_replication_iwinac09}, chip card verification
\cite{pretschner_et_al_model-based_testing_sttt04}, computational biology
\cite{bavarian_dahl_bio_seq_analysis_jucs06}, and decision support for cancer
diagnosis \cite{alma_thesis05}. Essentially, CHR performs guarded multiset
rewriting, extended by a complete and decidable constraint theory. A specific
strength of CHR is the wide array of available program analyses. Other formalisms
have been embedded in CHR in order to compare and mutually benefit from different
analysis approaches (cf. Section~\ref{sec:related_work}). In this work, we extend
this line of research by embedding graph transformation systems in CHR and
comparing confluence and operational equivalence analysis methods.

First, we embed graph transformation systems in CHR \cite{raiser07iclp} in
Section~\ref{sec:encoding}. This encoding is intuitive and offers a clear
one-to-one correspondence between GTS and CHR rules. Our proposed encoding
characterizes a subset of CHR that closely corresponds to graph transformation
systems, and furthermore we prove its soundness and completeness. Then, we show
that CHR is capable of expressing infinite numbers of graphs, which we will call
\emph{partial graphs}, and their transformations in a finite way, thus
facilitating program analysis.

\begin{figure} 
\centerline{
\xymatrix{
& \ar@{=>}[dl]_{r_1} \sigma \ar@{=>}[dr]^{r_2} &\\
\sigma_1 \ar@{=>}[dr]^{*} & & \ar@{=>}[dl]_{*} \sigma_2\\
& \sigma_1' \simeq \sigma_2' &
}}
\caption{Confluence Property for rules $r_1$ and $r_2$}
\label{fig:confluence}
\end{figure}
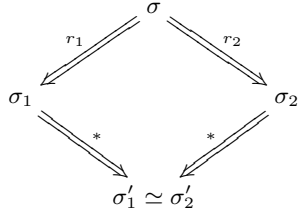

In non-deterministic rule-based systems, like GTS and CHR, two or more rules can
be applied to a state~$\sigma$. An interesting property in that respect is the
notion of \emph{confluence}, which holds, if for any case in which two rules are
applicable there exist computations yielding the same, or equivalent, results.
This situation is displayed in Figure~\ref{fig:confluence}, which due to its
shape is referred to as the \emph{diamond property}.

For terminating CHR programs a decidable automatic confluence test exists, based
on research in the area of term-rewriting \cite{baadernipkow98}. However as shown
in \cite{plump05}, an analogous approach fails for graph transformation systems.
Therefore, confluence analysis is an important example for a program analysis of
a GTS with methods from CHR. In Section~\ref{sec:confluence} we show that the
confluence test for CHR coincides with the strongest known sufficient criterion
for confluence of a GTS \cite{Raiser2009}.

\begin{figure} 
\centerline{
\xymatrix{
& \ar@{=>}[dl]_{\mcP_1} \sigma \ar@{=>}[dr]^{\mcP_2} &\\
\sigma_1 \ar@{=>}[dr]^{\mcP_1^*} & & \ar@{=>}[dl]_{\mcP_2^*} \sigma_2\\
& \sigma_1' \simeq \sigma_2' &
}}
\caption{Operational Equivalence for programs $\mcP_1$ and $\mcP_2$}
\label{fig:opeq}
\end{figure}
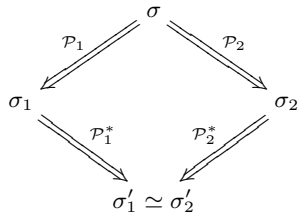

In Section~\ref{sec:opeq} we examine operational equivalence
\cite{abdennadherfruehwirth99} as a second example of a program analysis that is
available for CHR and can be applied to GTS. Operational equivalence,
intuitively, decides if two programs can compute equivalent results when given
the same input, as shown in Figure~\ref{fig:opeq}. The diamond shape in
Figure~\ref{fig:opeq} emphasizes the similarity to confluence, which is also
found in the respective program analysis methods. We introduce operational
equivalence in the GTS context in analogy to CHR \cite{Raiser2009b}. Then, we
prove that deciding operational equivalence of a CHR program, derived from a GTS,
is a sufficient criterion for operational equivalence of the corresponding GTS.

An interesting application of this result is the possibility to detect and
remove redundant rules using the test for operational equivalence. Redundant rules of
graph transformation systems have been formally defined in \cite{Kreowski2000},
however to the best of our knowledge, this is the first available algorithm for
detecting them in a GTS.

This work presents a unified treatment and considerable extension of previously
published works
\cite{raiser07iclp,Raiser2009a,Raiser2009c,Raiser2009b,Raiser2009}. In
\cite{Raiser2009a} a formal treatment of CHR state equivalence is provided and,
derived from that, a simplified formulation of the operational semantics of CHR.
This novel formulation allows us to unify our previous works while simplifying
presentation and formal proofs significantly. Furthermore, the state equivalence
definition from \cite{Raiser2009a} is the basis for new insights on CHR states
that encode graphs.

\section{Preliminaries}
\label{sec:preliminaries}

In this section we introduce the required formalisms for graph transformation
systems in Section~\ref{sec:prelim:gts} and constraint handling rules in
Section~\ref{sec:prelim:chr}.

\subsection{Graph Transformation System}
\label{sec:prelim:gts}

The following definitions for graphs and graph transformation systems (GTS) have
been adapted from \cite{ehrigprangetaentzer06}.

\begin{definition}[graph] A \emph{graph} $G = (V, E, \src,
\tgt)$ consists of a finite set $V$ of nodes, a finite set $E$ of edges and two
functions $\src, \tgt: E \rightarrow V$ specifying source and target of an edge,
respectively. A \emph{type graph} $TG$ is a graph with unique labels for all
nodes and edges.

For simplicity, we avoid an additional label function in favor of identifying
variable names with labels. For multiple graphs we refer to the node~set~$V$ of a
graph~$G$ as $V_G$ and analogously for edge sets and the $\src, \tgt$ functions.
We further define the degree of a node as $\deg : V \rightarrow \mathbb{N}, v
\mapsto \#\{ e \in E \mid \src(e) = v \} + \#\{ e \in E \mid \tgt(e) = v \}$. As
there may be multiple graphs containing the same node, we use $\deg_G(v)$ to
specify the degree of a node~$v$ with respect to the graph~$G$. When the context
graph is clear the subscript is omitted.
\end{definition}

In this work, we consider typed graphs, i.e. graphs in which nodes and edges are
assigned types from a type graph.

\begin{definition}[graph morphism,typed graph] 
Given graphs $G_1, G_2$ with $G_i = (V_i,E_i,\src_i,\tgt_i)$ for $i=1,2$ a
\emph{graph morphism} $f: G_1 \rightarrow G_2, f = (f_V,f_E)$ consists of two
functions $f_V: V_1 \rightarrow V_2$ and $f_E : E_1 \rightarrow E_2$ that
preserve the source target functions, i.e. $f_V \circ \src_1 = \src_2 \circ
f_E$ and $f_V \circ \tgt_1 = \tgt_2 \circ f_E$.

A graph morphism $f$ is \emph{injective} (or \emph{surjective}) if both
functions~$f_V,f_E$ are injective (or surjective, respectively); $f$ is called
\emph{isomorphic} if it is bijective. $f$ is called an \emph{inclusion} if
$f_V(V_1) \subseteq V_1$ and $f_E(E_1) \subseteq E_1$. When the context is
clear, we simply refer to graph morphisms as morphisms.

A \emph{typed graph} $G$ is a tuple $(V, E, \src, \tgt, \type, TG)$ where $(V, E,
\src, \tgt)$ is a graph, $TG$ a type graph, and $\type$ a graph morphism with
$\type = (\type_V, \type_E)$ and $\type_V : V \rightarrow TG_V, \type_E : E
\rightarrow TG_E$.

For a typed graph~$G = (V, E, \src, \tgt, \type, TG)$ we define a
\emph{subgraph}~$H$ as a typed graph~$(V',E',\src',\tgt',\type',TG)$ such that
$V' \subseteq V \land E' \subseteq E \land \src' = \src \mid_{E'} \land \tgt' =
\tgt \mid_{E'} \land \type_V' = \type_V \mid_{V'} \land \type_E' = \type_E
\mid_{E'}$ with $\forall e \in E' . \src'(e) \in V' \land \tgt'(e) \in V'$.
\end{definition}

\begin{figure}
\centerline{
\scalebox{0.4}{\includegraphics{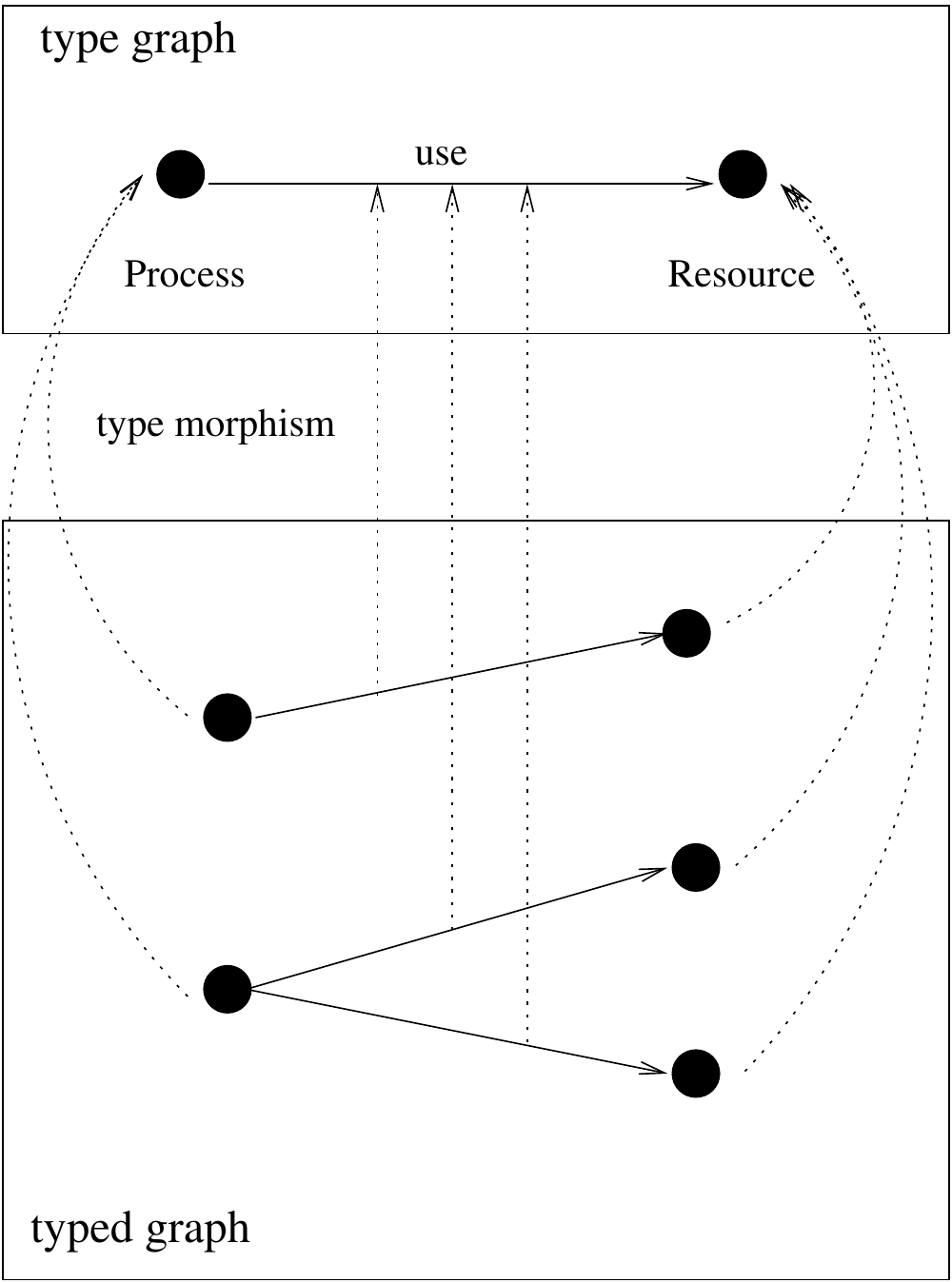}}
}
\caption{Example of a type graph and typed graph}
\label{fig:type}
\end{figure}

\begin{example}Figure~\ref{fig:type} shows an example for a type graph and a
corresponding typed graph. The type graph at the top defines two types of nodes:
processes and resources. Furthermore, it defines \emph{use} edges going from
processes to resources. The typed graph is one possible instance of a graph
modeling processes and resources being used by those processes. The $\type$
graph morphism is represented by the dotted lines, showing how the nodes are
typed as processes or resources, respectively.
\end{example}

\begin{definition}[GTS, rule] A \emph{Graph Transformation System} (GTS) is a
tuple consisting of a type graph and a set of graph production rules. A
\emph{graph production rule} -- simply called \emph{rule} if the context is clear
-- is a tuple \prodrule of graphs $L,K$, and $R$ with inclusion morphisms $l:K
\rightarrow L$ and $r:K \rightarrow R$.
\end{definition}

We distinguish two kinds of typed graphs: \emph{rule graphs} and \emph{host
graphs}. Rule graphs are the graphs $L, K, R$ of a graph production rule $p$ and
host graphs are graphs to which the graph production rules are applied. This work
is based on the double-pushout approach (DPO) as defined in
\cite{ehrigprangetaentzer06}. Most notably, we require a \emph{match~morphism}~$m
: L \rightarrow G$ to apply a rule $p$ to a typed host graph $G$. The
transformation yielding the typed graph $H$ is written as $G
\stackrel{p,m}{\Longrightarrow} H$. $H$ is given mathematically by constructing
$D$ as shown in Figure~\ref{fig:dpo}, such that (1) and (2) are pushouts in the
category of typed graphs. Intuitively, the graph~$L$ is matched to a subgraph of
$G$ and its occurrence in $G$ is then replaced by the graph~$R$. The intermediate
graph~$K$ is the \emph{context graph}, which contains the nodes and edges in
both $L$ and $R$, i.e. all nodes and edges matched to $K$ remain during the
transformation.

\begin{figure} 
 \centerline{
  \xymatrix { 
   L \ar[d]_m \ar @{} [dr]|{(1)} & \ar[l]_l \ar[d]_k K \ar[r]^r \ar @{}
   [dr]|{(2)} & \ar[d]_n R \\ G  & \ar[l]_f D \ar[r]^g & H
  }
 }
  
 \caption{Double-pushout approach}
 \label{fig:dpo}
\end{figure}

A graph production rule $p$ can only be applied to a host graph $G$ if the
following gluing condition is satisfied. In fact, \cite{ehrigprangetaentzer06}
shows, that $D$ and the pushout~(1) exist if and only if this gluing condition is
satisfied. It is based on the following three sets \cite{ehrigprangetaentzer06}:

\begin{itemize}
  \item gluing points: $GP = l(K)$
  \item identification points: $IP = \{ v \in V_L \mid \exists w \in V_L, w \ne 
  v : m(v)=m(w) \} \cup \{ e \in E_L \mid \exists f \in E_L, e \ne f : 
  m(e)=m(f) \}$
  \item dangling points: $DP = \{ v \in V_L \mid \exists e \in E_G \setminus 
  m(E_L) : \src_G(e) = m(v) \vee \tgt_G(e)=m(v) \}$
\end{itemize}

\begin{definition}[gluing condition]
The \emph{gluing condition} is defined as $IP \cup DP \subseteq GP$.
\end{definition}

If the gluing condition is satisfied for a rule $\prodrule$ the application of
the rule consists of transforming $G$ into $H$ by performing the construction
described above. An implementation-oriented interpretation of a rule application
is that all nodes and edges in $m(L \setminus l(K))$ are removed from $G$ to
create $D = (G \setminus m(L)) \cup m(l(K))$ and then all nodes and edges in $n(R
\setminus r(K))$ are added to create $H = D \cup n(R \setminus r(K))$.

\begin{example}\label{ex:gts}
Figure~\ref{fig:cyclic_grs} shows two graph production rules in a shorthand
notation that defines the morphisms $l$ and $r$ implicitly by the labels of the
nodes which are mapped onto each other. The resulting graph transformation system
is implicitly defined over the simple type graph consisting only of a single node
with a loop, depicted in Figure~\ref{fig:trivial_type}.

The two rules constitute a graph transformation system for detecting cyclic
lists. The basic idea of the \emph{unlink} rule is to remove intermediate nodes
of the list, while the \emph{twoloop} rule replaces the cyclic list consisting of
two nodes by a single node with a loop. Note that application of the
\emph{twoloop} rule requires that no additional edges are adjacent to the removed
node. Such \emph{dangling edges} are discussed in more detail in
Section~\ref{sec:encoding}.

To detect if a host graph is a cyclic list, the GTS is applied to the host graph
until exhaustion, i.e. until no rule is applicable anymore. The initial host
graph then is a cyclic list if and only if the final graph consists of a single
node with a loop (cf. \cite{bakewellplumprunciman03}).

\begin{figure}
\centerline{
\scalebox{0.5}{\includegraphics{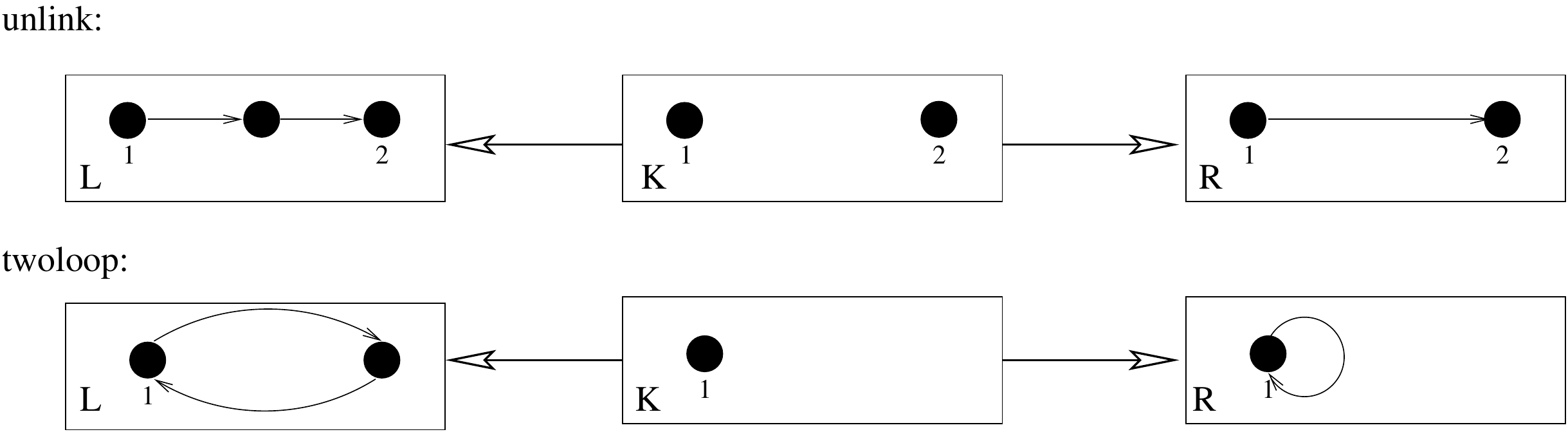}}
}
\caption{Graph transformation system for recognizing cyclic lists}
\label{fig:cyclic_grs}
\end{figure}

\begin{figure}
\includegraphics{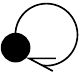} 
\caption{Simple type graph consisting of a node and edge}
\label{fig:trivial_type}
\end{figure}
\end{example}

In general, the match morphism $m$ can be non-injective. However, for the
remainder of this work we only consider injective match morphisms, which have the
advantage that the set $IP$ of identification points is guaranteed to be
$\emptyset$. Furthermore, non-injective match morphisms can be simulated as
follows: given a rule $\prodrule$ and a non-injective match morphism $m$ it holds
$\forall v,w \in V_L, v \ne w$ with $m(v) = m(w)$ that the rule is only
applicable, if $v,w \in l(V_K)$, i.e. only nodes which are not removed by the
rule application are allowed to be matched non-injectively -- otherwise $IP \not
\subseteq GP$. Therefore, it is possible to add another rule $p'$ which is
derived from $p$ by merging the nodes $v$ and $w$ into a node $v_w$ in all three
graphs of the rule. Thus, the non-injective matching with $m(v) = m(w)$ can be
simulated by injectively matching $v_w$ to $m(v_w)$ where $m(v_w)$ is the same
node in $G$ as $m(v)$. The same argumentation holds for edges, analogously.
Therefore, we can restrict ourselves to injective match morphisms by extending
the set of rules with new rules for all possible merges of nodes and edges in the
graph $K$. This simplifies the generic gluing condition to $DP \subseteq GP$.

Finally, we require the following definition of the track morphism
\cite{Plump1995}. Intuitively, the track morphism is defined for a node or edge,
if it is not removed by the rule application.

\begin{definition}[track morphism]\label{def:track} Given $G \dergts H$
the \emph{track morphism} $\track_{G \dergts H}: G \rightarrow H$ is the partial
graph morphism defined by $$ \track_{G \dergts H}(x) = \left\{
 \begin{matrix}
 g(f^{-1}(x)) & \text{ if } x \in f(D),\\
 \text{undefined} & \text{otherwise.}
 \end{matrix} \right.
$$
Here $f: D \rightarrow G$ and $g : D \rightarrow H$ are the morphisms in the
lower row of the pushout~(1) in Figure~\ref{fig:dpo} and $f^{-1} : f(D)
\rightarrow D$ maps each item $f(x)$ to $x$.

The track morphism of a derivation $\Delta : G_0 \dergts^* G_n$ is defined by
$\track_{\Delta} = \id_{G_0}$ if $n = 0$ and $\track_{\Delta} = \track_{G_1
\dergts^* G_n} \circ \track_{G_0 \dergts G_1}$ otherwise, where
$\id_{G_0}$ is the identity morphism on $G_0$.
\end{definition}

\subsection{Constraint Handling Rules}
\label{sec:prelim:chr}
This section presents the syntax and operational semantics of Constraint Handling
Rules (CHR) \cite{chr_survey_tplp08,fruehwirth09}. Constraints are first-order
predicates which we separate into \emph{built-in constraints} and
\emph{user-defined constraints}. Built-in constraints are provided by the
constraint solver while user-defined constraints are defined by a CHR program.
The notation~$c/n$, where $c$ is called the \emph{constraint symbol} and $n$ the
\emph{arity}, is used for both types of constraints.

Its semantics is based on an underlying complete \emph{constraint theory}~$\CT$
on built-in constraints for which satisfiability and entailment are decidable. In
general, CHR allows arbitrary constraint theories for \CT, requiring only that it
contains at least Clark's equality theory for syntactic equality. In addition to
that, in this work we also require \CT\ to cover the elementary arithmetic
operations~${+}$ and ${-}$. Furthermore, $\top$ denotes the built-in which is
always true and $\bot$ denotes false, respectively.

The survey \cite{chr_survey_tplp08} provides an overview over the different
techniques used in CHR implementations and the book \cite{fruehwirth09} details
the different available operational semantics for CHR. In this work we abstract
from specific implementations and rely on the operational semantics given in
\cite{Raiser2009a}, which corresponds to the \emph{very abstract operational
semantics} in \cite{fruehwirth09}.

CHR is a state transition system over the set of states given in the following
definition.

\begin{definition}[CHR states] 
\label{def:states}

A \emph{(CHR) state} is a tuple \[\st{\bbG}{\bbB}{\bbV}.\]

\bbG\ is a multiset of user-defined constraints called the \emph{goal} (or
\emph{(user-defined) constraint store}), \bbB\ is a conjunction of built-in
constraints called the \emph{built-in (constraint) store}, and \bbV\ is the set
of \emph{global variables}.

In this work $\sigma, \tau, \ldots$ denote CHR states and $\Sigma$ denotes the
set of all CHR states.
\end{definition}

The following definition introduces the different types of variables we
distinguish for a given CHR state.

\begin{definition}[Variable Types]\label{def:var_types} For the variables
occurring in a state~$\sigma = \st{\bbG}{\bbB}{\bbV}$ we distinguish three
different types:
\begin{enumerate}
  \item a variable~$v \in \bbV$ is called a \emph{global} variable
  \item a variable~$v \not \in \bbV$ is called a \emph{local} variable
  \item a variable~$v \not \in (\bbV \cup \vars(\bbG))$ is called a
  \emph{strictly local} variable
\end{enumerate}
\end{definition}

The following equivalence relation~$\equiv$ between CHR states \cite{Raiser2009a}
is an important tool that facilitates a succinct definition of the operational
semantics of CHR and simplifies proofs.

\begin{definition}[State Equivalence]
\label{def:equiv}

Equivalence between CHR states is the smallest equivalence relation~$\equiv$
over CHR states that satisfies the following conditions:

\begin{enumerate}
\item \label{cond:subst} \emph{(Substitution)}
\[
	\st{\bbG}{x \doteq t \land \bbB}{\bbV} \equiv
	\st{\bbG\subxt}{x \doteq t \land \bbB}{\bbV}
\]
\item \label{cond:ct} \emph{(Transformation of the Constraint Store)} If
$\CT\models\exists \bar s.\bbB \leftrightarrow\exists\bar s'.\bbB'$ where $\bar
s, \bar s'$ are the strictly local variables of $\bbB,\bbB'$, respectively,
then:
\[
	\st{\bbG}{\bbB}{\bbV} \equiv \st{\bbG}{\bbB'}{\bbV}
\]
\item \label{cond:global} \emph{(Omission of Non-Occurring Global Variables)} If
$X$ is a variable that does not occur in $\bbG$ or $\bbB$ then:
\[
	\st{\bbG}{\bbB}{\{X\}\cup\bbV} \equiv \st{\bbG}{\bbB}{\bbV}
\]
\item \label{cond:fail} \emph{(Equivalence of Failed States)}
\[
	\st{\bbG}{\bot}{\bbV} \equiv \st{\bbG'}{\bot}{\bbV}
\]
\end{enumerate}

\end{definition}

The following lemma presents basic properties of this equivalence relation:

\begin{lemma}[Properties of State Equivalence \cite{Raiser2009a}]
\label{lem:equiv_props}
The equivalence relation over CHR states, given in Definition~\ref{def:equiv},
has the following properties:

\begin{enumerate}
\item \label{prop:rename} \emph{(Renaming of Local Variables)} Let $x,y$ be
variables such that $x,y\not\in\bbV$ and $y$ does not occur in $\bbG$ or $\bbB$:
\[
	\st{\bbG}{\bbB}{\bbV} \equiv \st{\bbG\subxy}{\bbB\subxy}{\bbV}
\]
\item \label{prop:partial} \emph{(Partial Substitution)} Let $\bbG\psubxt$ be a
multiset where \emph{some} occurrences of $x$ are substituted with $t$:
\[
	\st{\bbG}{x \doteq t \land \bbB}{\bbV} \equiv
	\st{\bbG\psubxt}{x \doteq t \land \bbB}{\bbV}
\]
\item \label{prop:lequiv} \emph{(Logical Equivalence)} If
\[
	\st{\bbG}{\bbB}{\bbV} \equiv \st{\bbG'}{\bbB'}{\bbV'}
\]
then $\CT \models \exists \bar y.\bbG \land \bbB \leftrightarrow
\exists \bar y'.\bbG' \land \bbB'$,  where $\bar y,\bar y'$ are the local
variables of $\st{\bbG}{\bbB}{\bbV},  \st{\bbG'}{\bbB'}{\bbV'}$, respectively.
\end{enumerate}
\end{lemma}

Decidability of state equivalence is a result of the following theorem from
\cite{Raiser2009a}:

\begin{theorem}[Criterion for $\equiv$ \cite{Raiser2009a}]\label{thm:equiv_tf} Let
$\sigma = \st{\bbG}{\bbB}{\bbV}, \sigma' = \st{\bbG'}{\bbB'}{\bbV}$ be CHR states with
local variables~$\bar y,\bar y'$ that have been renamed apart.\[ \sigma \equiv
\sigma' \text{ iff } \CT \models \forall (\bbB \rightarrow \exists
\bar y'. ((\bbG = \bbG') \land \bbB')) \land \forall (\bbB' \rightarrow \exists
\bar y.((\bbG = \bbG') \land \bbB)) \]
\end{theorem}

As CHR is a rule-based programming language we now introduce the different
types of possible CHR rules.

\begin{definition}[CHR Rules, CHR Program]\label{def:rules}
For multisets~$H_1,H_2,B_c$ of user-defined constraints with $H_1, H_2
\ne \emptyset$ and conjunctions $G,B_b$ of built-in constraints a CHR
\emph{simpagation} rule is of the form
\[
H_1 \backslash H_2 \Leftrightarrow G \mid B_c,B_b.
\]
For the case~$H_1 = \emptyset$ we call the rule a \emph{simplification} rule
and denote it as
\[
H_2 \Leftrightarrow G \mid B_c,B_b
\]
and for the case~$H_2 = \emptyset$ we call the rule a \emph{propagation} rule
and denote it as
\[
H_1 \Rightarrow G \mid B_c,B_b.
\]
If $G = \top$ it can be omitted together with the $'\mid'$.

A \emph{CHR program} is a set of CHR rules.
\end{definition}

Next, we define the operational semantics of CHR by introducing its transition
relation~$\der$ based on the formulation given in \cite{Raiser2009a}, which
relies on equivalence classes of CHR states. In the remainder of this work we
take the liberty of notationally identifying a CHR state~$\sigma$ with its
equivalence class~$[\sigma]$. Furthermore, we simplify multiset expressions like
$\{a\} \uplus \{b\}$ to $a \uplus b$ or $a,b$.

\begin{definition}[Operational Semantics] \label{def:opsem}

For a CHR program~$\mcP$ we define the state transition
system~$(\Sigma/\!\!\equiv, \der)$ as follows. The application of a rule~$r \in
\mcP$ assumes a copy of it that contains only fresh variables.

\begin{center}
\textwidth 9.5cm
\begin{tabular}{c}
$r\ @\ H_1 \backslash H_2 \Leftrightarrow G\mid B_c, B_b$ \\
\hline
$[\st{H_1 \uplus H_2 \uplus \bbG}{G \land \bbB}{\bbV}]
	\der
[\st{H_1 \uplus B_c \uplus \bbG}{G \land B_b \land \bbB}{\bbV}]$
\end{tabular}
\end{center}
\end{definition}

Simplification rules are only syntactically different, but operate as described
by Definition~\ref{def:opsem} with $H_1 = \emptyset$, respectively. Note that
propagation rules lead to trivial non-termination in this formulation, however
that is no problem here, because the work at hand requires no propagation rules.

A rule~$r \in \mcP$ is \emph{applicable} to a state~$\sigma$ if and only if there
exists a state~$\tau$ such that $\sigma \der \tau$. We say that a state~$\sigma$
is \emph{final} if and only if there exists no state~$\tau$ with $\sigma \der
\tau$. As usual, $\der^*$ denotes the reflexive-transitive closure of $\der$.
When we want to emphasize that a transition uses a specific rule~$r$ we denote
this by $\der^r$. When discussing multiple programs, $\der_\mcP$ denotes a
transition using a rule of program~\mcP.

\begin{example}[Example Computation]
In this comprehensive example, we present a complete computation in CHR. Readers
already familiar with CHR may want to skip this.

The following rule \cite{fruehwirth09} is a program for computing the minimum of
a multiset of numbers: \[ \min(N) \backslash \min(M) \Leftrightarrow N \le M
\mid \top \]

Intuitively, two $\min$ constraints are matched and the one with the larger
number is removed. We will now walk through the detailed computation of running
the following input~$\sigma$ on the above program, in order to determine the
minimum of the numbers~$1,3$, and $4$: \[ \sigma = \st{\min(1) \uplus \min(3)
\uplus \min(X)}{X=4}{\{X\}} \]

First, we take a fresh copy of the rule as demanded by
Definition~\ref{def:opsem}: \[ \min(N_1) \backslash \min(M_1) \Leftrightarrow
N_1 \le M_1 \mid \top \]

Next, we apply Definition~\ref{def:equiv} in order to show that $\sigma$ is
contained in the equivalence class required for applying this rule (we use
$\bbV = \{X\}$ here):
\[
\begin{array}{cl}
\sigma \eqct & \st{\min(1) \uplus \min(3) \uplus \min(X)}{N_1 \le M_1 \land X=4
\land N_1=1 \land M_1=3}{\bbV} \\
\eqsubst & \st{\min(N_1) \uplus \min(M_1) \uplus \min(X)}{N_1 \le M_1 \land
X=4 \land N_1=1 \land M_1=3}{\bbV}\\
= & \st{\min(N_1) \uplus \min(M_1) \uplus \bbG}{N_1 \le M_1 \land \bbB}{\bbV}
\end{array}
\]

Hence, all conditions for Definition~\ref{def:opsem} are satisfied, so we can
apply the rule to the equivalence class of $\sigma$, getting $\sigma \der
\tau$, or more precisely, $[\sigma] \der [\tau]$: 
\[ 
\begin{array}{cl}
\sigma \der & \st{\min(N_1) \uplus \bbG}{N_1 \le M_1 \land \top \land
\bbB}{\bbV}\\
= & \st{\min(N_1) \uplus \min(X)}{N_1 \le M_1 \land \top \land X=4 \land N_1 = 1
\land M_1 = 3}{\bbV} \\
\eqsubst & \st{\min(1) \uplus \min(X)}{N_1 \le M_1 \land \top \land X=4 \land
N_1=1 \land M_1 = 3}{\bbV} \\
\eqct & \st{\min(1) \uplus \min(X)}{X=4}{\bbV} = \tau
\end{array}
\]

Next, we repeat this procedure for another application of the above rule, based
on the following fresh copy: \[ \min(N_2) \backslash \min(M_2) \Leftrightarrow
N_2 \le M_2 \mid \top \]

This results in the expected answer, that $1$ is the minimum of the
numbers~$1,3$, and $4$:
\[
\begin{array}{cl}
\tau \eqct & \st{\min(1) \uplus \min(X)}{N_2 \le M_2 \land N_2 = 1 \land M_2
= X \land X = 4}{\bbV} \\
\eqsubst & \st{\min(N_2) \uplus \min(M_2)}{N_2 \le M_2 \land N_2 = 1 \land
M_2 = X \land X =4 }{\bbV}\\
\der & \st{\min(N_2)}{N_2 \le M_2 \land \top \land N_2 = 1 \land M_2 = X
\land X = 4}{\bbV} \\
\eqsubst & \st{\min(1)}{N_2 \le M_2 \land \top \land N_2 = 1 \land M_2 = X
\land X=4}{\bbV}\\
\eqct & \st{\min(1)}{X=4}{\bbV}
\end{array}
\]

We can also witness the difference between global and local variables in this
computation. While the variable~$X$ is no longer used in a CHR constraint in the
final state, we still have to keep track of the information~$X=4$, because it is
a global variable. The auxiliary variables~$N_1,M_1,\ldots$ instead, are local
when used in a CHR constraint and strictly local, when only occurring in the
built-in store. In the latter case we may replace the built-in store by a
logically equivalent representation that removes the strictly local variables.
\end{example}

\section{Embedding GTS in CHR}
\label{sec:encoding}

In this section we encode rules of a graph transformation system as CHR rules and
discuss how host graphs are encoded in CHR to work with these rules.
Section~\ref{sec:default_encoding}  defines the necessary encoding and presents
an example computation in CHR. We then analyze formal properties of graph
transformation systems embedded in CHR in Section~\ref{sec:properties}. Finally,
Section~\ref{sec:encoding_discussion} discusses the suitability of this encoding
for program analysis and variations of the encoding.

In this work, we assume that the CHR programs resulting from encoding a GTS are
executed only with encodings of graphs. Naturally, we may provide the CHR
programs with completely different inputs or inconsistently encoded graphs. It is
clear, that we cannot expect any meaningful results from such computations,
hence, for the remainder of this work we restrict all observations to programs
and states that correspond to GTS and graphs. We formalize this restriction in
Section~\ref{sec:properties} by means of an invariant. Therefore, on one hand any
state that violates the invariant will not be considered as input, and on the
other hand any graph can be encoded into a state that satisfies the invariant. We
show in Section~\ref{sec:sound_complete} that execution of the encoded GTS in CHR
for invariant-satisying states always leads to results that also satisfy the
invariant. In other words, when providing a graph as input to the CHR program,
the result will also be a graph, as is to be expected.

\subsection{CHR Encoding of a GTS}
\label{sec:default_encoding}

First, we determine the necessary constraint symbols for encoding rule and host
graphs. At this point we require the GTS to be typed, so this can be directly
inferred from the corresponding type graph as explained in
Definition~\ref{def:constraints_i}. Note that this is not a restriction though,
as every untyped graph can be typed over the type graph consisting of a single
node with a loop (cf. Figure~\ref{fig:trivial_type}).

\begin{definition}[type graph encoding] \label{def:constraints_i}
For a type graph $TG$ we define the set $\mathcal{C}$ of required constraint
symbols to encode graphs typed over $TG$ as the minimal set satisfying:
\begin{itemize}
  \item If $v \in V_{TG}$ then $v/2 \in \mathcal{C}$.
  \item If $e \in E_{TG}$ then $e/3 \in \mathcal{C}$.
\end{itemize}
\end{definition}

We assume that all constraints introduced by Definition~\ref{def:constraints_i}
have unique names. Furthermore, for graphs to be encoded with these constraints,
we associate elements of the set~$V$ of nodes with integer numbers or letters
that can be used as arguments.

\begin{definition}[typed graph encoding]\label{def:encoding}
 
We define the following helpful mappings for an infinite set
of variables~\VARS:

\begin{itemize}
  \item $\type_G(x)$ denotes the corresponding constraint symbol for encoding a
  node or edge of the given type.

  \item $\var : G \rightarrow \VARS, x \mapsto X_x$ such that $X_x$ is a unique
  variable associated to $x$, i.e. $\var$ is injective for $X$ being the set of
  all graph nodes and edges.
  
  \item $\dvar : G \rightarrow \VARS, x \mapsto X_x$ such that $X_x$ is a unique
  variable associated to $x$, i.e. $\dvar$ is injective for $X$ being the set of
  all graph nodes and edges and different from $\var$.
\end{itemize}

Using these mappings we define the following encoding of graphs:
\[
\chr_G(E, x) = \left\{ \begin{array}{lp{3cm}}
	\type_G(x)( \var(x), \deg_G(x)) & \hskip-1cm\text{ if } $x \in V_G \land
	E=\gnd$ \\ \type_G(x)( \var(x), \dvar(x)) & \hskip-1cm\text{ if } $x \in V_G
	\land E =\kp$\\ \type_G(x)( \var(x), \var(\src(x)), \var(\tgt(x))) & \text{ if } $x \in
	E_G$ \end{array}\right.
\]

We use the notations $\chr(\gnd, G) = \{ \chr_G(\gnd, x) \mid x \in G \}$ as
well as $\chr(\kp, G) = \{ \chr_G(\kp, x) \mid x \in G \}$. Furthermore, we omit the
index~$G$ if the context is clear. We call $\dvar(v)$ the \emph{degree variable}
for a node~$v$.

A host graph~$G$ is encoded in CHR as $\st{\chr(\gnd,G)}{\top}{\bbV}$, where
\bbV\ can be chosen freely.
\end{definition}

\begin{figure}
\includegraphics{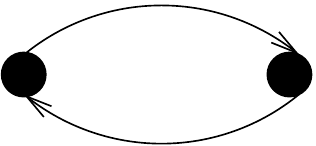}
\caption{Cyclic graph consisting of two nodes} 
\label{fig:loop2}
\end{figure}

\begin{example}[cont] For our example of the GTS for recognizing cyclic lists we
assume the type graph in Figure~\ref{fig:trivial_type}. Based on this type graph
we need the constraints $\node/2$ and $\edge/3$. The host graph~$G$ given in
Figure~\ref{fig:loop2} that contains a cyclic list consisting of exactly two
nodes is encoded in $\chr(\gnd, G)$ as: \[ \node(N_1, 2),
\node(N_2, 2), \edge(E_1, N_1, N_2), \edge(E_2, N_2, N_1) \] The same
graph~$G$ encoded in $\chr(\kp, G)$ has the following form: \[ \node(N_1, D_1),
\node(N_2, D_2), \edge(E_1, N_1, N_2), \edge(E_2, N_2, N_1) \]
\end{example}

We can now encode a complete graph production rule based on these definitions:

\begin{definition}[GTS rule in CHR]
\label{def:chr_rule}
For a graph production rule~\prodrule from a GTS we define $\chrrule(p) =
(p\ @\ C_L \Leftrightarrow C_R^u, C_R^b)$ with
\begin{itemize}
  \item $C_L = \{ \chr_L(\kp, x) \mid x \in K \} \uplus \{ \chr_L(\gnd, x) \mid
  x \in L \setminus K \}$
  \item $C_R^u = \{ \chr_R(\gnd, x) \mid x \in R \setminus K\} \uplus
  \{\chr_R(\kp, e) \mid e \in E_K \}$\\ $\uplus \{ \chr_R(\kp, v') \mid v
  \in V_K\}$
  \item $C_R^b = \{ \var(v) = \var(v') \land \dvar(v') =
  \dvar(v){-}\deg_L(v){+}\deg_R(v) \mid v \in V_K \}$
\end{itemize}
\end{definition}

A CHR program that is created from a GTS according to the above definition, will
be referred to as a \emph{GTS-CHR program} for the remainder of this work.

\begin{example}[cont.]\label{ex:twoloop}As an example, consider the second rule
from Example~\ref{ex:gts}, which reduces two cyclic nodes to a single node with
a loop. Its encoding as a CHR simplification rule is given below: \\
\begin{tabular}{ll} 
twoloop $@$ & $\node(N_1, D_1) \uplus \node(N_2, 2) \uplus$\\
& $\edge(E_1, N_1, N_2) \uplus \edge(E_2, N_2, N_1)$\\
& $\Leftrightarrow$\\
& $\node(N_1', D_1') \uplus \edge(E_3, N_1, N_1), N_1' = N_1 \land D_1' =
D_1{-}2{+}2$
\end{tabular}

Note that it is also possible to simplify this encoding, as explained later in
Section~\ref{sec:diff_encoding}.
\end{example}

When applying a GTS rule the gluing condition has to be satisfied. Due to our
restriction to injective match morphisms, the gluing condition is violated if
there exists $x \in DP$ with $x \not \in GP$. Intuitively, when a node gets
deleted by a rule, the corresponding node in the host graph may have an edge
adjacent to it which is not explicitly given in the rule. In such a case, the
remaining edge would be left \emph{dangling} as it is no longer adjacent to two
nodes. Therefore, this situation has to be avoided and before a rule is applied
to a host graph, we first have to ensure that there are no dangling edges
according to the following definition:

\begin{definition}[dangling edge]A \emph{dangling edge} is an edge $e \in E_G 
\setminus m(E_L)$ such that there is a node $v \in V_L \setminus V_K$ with 
$m(v) = \src(e) \lor m(v) = \tgt(e)$.
\end{definition}

\begin{figure}
\includegraphics{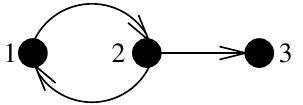}
\caption{Graph with a dangling edge if node~2 is removed by the
\emph{twoloop} rule}
\label{fig:dangle}
\end{figure}

\begin{example}[cont.] Consider the \emph{twoloop} rule given in
Example~\ref{ex:twoloop}, along with the following encoded host graph shown in
Figure~\ref{fig:dangle}:
\[
\begin{array}{l}
\node(V_1, 2), \node(V_2, 3), \node(V_3,1),\\
\edge(E_1, V_1, V_2), edge(E_2, V_2, V_1), \edge(E_3, V_2, V_3)
\end{array}
\]

Applying the \emph{twoloop} rule to this graph to remove the node $V_2$ would
leave the edge~$E_3$ dangling. However, this is avoided as the encoding of the
\emph{twoloop} rule contains the following constraint in its head: $\node(N_2,
2)$. Hence, only a node with a degree of exactly $2$ can be removed by this rule.
Nevertheless, the rule can be applied with $N_2 = V_1$ as the node $V_1$ has the
required degree of $2$.
\end{example}

\subsubsection{Example Computation}
\label{sec:encoding_example}

In this section we provide a complete computation for our cyclic list example to
demonstrate how our encoding works. The following two rules are the CHR encoding
of the rules in Figure~\ref{fig:cyclic_grs}:

\begin{center} 
\begin{tabular}{ll}
\emph{unlink}\ @ & $\node(N_1, D_1) \uplus \node(N, 2) \uplus \node(N_2,
D_2) \uplus$\\
				& $\edge(E_1, N_1, N) \uplus \edge(E_2, N, N_2)$\\
                & $\Leftrightarrow$\\
                & $\node(N_1', D_1') \uplus \node(N_2', D_2') \uplus \edge(E,
                N_1, N_2),$\\
                & $N_1' = N_1 \land N_2' = N_2 \land D_1' = D_1{+}1{-}1
                \land D_2' = D_2{+}1{-}1$\\
                &\\
\emph{twoloop}\ @ & $\node(N_1, D_1) \uplus \node(N, 2) \uplus$ \\
				 & $\edge(E_1, N_1, N) \uplus \edge(E_2, N, N_1)$ \\
				 & $\Leftrightarrow$ \\
				 & $\node(N_1', D_1') \uplus \edge(E, N_1, N_1),$\\
				 & $N_1' = N_1 \land D_1' = D_1{+}2{-}2$
\end{tabular}
\end{center}

\begin{figure}
\includegraphics{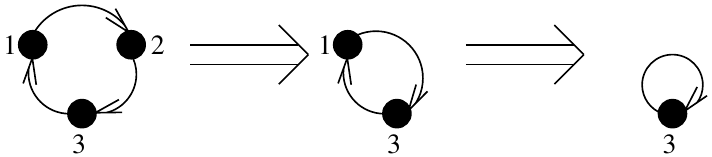}
\caption{Example computation} 
\label{fig:ex_comp}
\end{figure}

The following state~$\sigma$ encodes a cycle consisting of three nodes. The
following computation is depicted in Figure~\ref{fig:ex_comp}. To demonstrate
computations on partially defined graphs, further discussed in
Section~\ref{sec:encoding_discussion}, the degree of the third node is left
uninstantiated:

$\sigma = \st{\node(N_1, 2) \uplus \node(N_2, 2) \uplus \node(N_3, D_3) \uplus\\
\edge(E_1, N_1, N_2) \uplus \edge(E_2, N_2, N_3) \uplus \edge(E_3,
N_3, N_1)}{\\\top}{\{N_1, N_2, N_3, E_1, E_2, E_3, D_3\}}$

Rule~\emph{unlink} is applied to state~$\sigma$ resulting in the state

$\st{\node(N_1', D_1') \uplus \node(N_3', D_3') \uplus \edge(E, N_1,
N_3) \uplus \edge(E_3, N_3, N_1)}{\\N_1' = N_1 \land D_1' = 2{+}1{-}1
\land N_3' = N_3 \land D_3' = D_3{+}1{-}1}{\{N_1, N_2, N_3, E_1, E_2, E_3,
D_3\}}$

which is equivalent to state~$\sigma'$:

$\sigma' = \st{\node(N_1, 2)\uplus \node(N_3, D_3) \uplus \edge(E, N_1,
N_3) \uplus \edge(E_3, N_3, N_1)}{\\\top}{\{N_1, N_3, E_3, D_3\}}$

Finally, rule~\emph{twoloop} is applied to $\sigma'$ to remove node~$N_1$,
resulting in $\sigma''$:

$\sigma'' = \st{\node(N_3, D_3) \uplus \edge(E', N_3, N_3)}{\top}{\{N_3, D_3\}}$

As can be seen the built-in store may contain a chain of degree adjustments for
nodes with initially uninstantiated degree, although in this example it is not
the case as all degrees remain unchanged. The other interesting consequences of
partially uninstantiated encodings are investigated more thoroughly in
Section~\ref{sec:encoding_discussion}.

\subsection{Formal Properties}
\label{sec:properties}

This section examines formal properties of the encoding given in
Section~\ref{sec:default_encoding}. First, Section~\ref{sec:graph_states}
analyzes the special CHR states found when working with a GTS-CHR program. Then
we prove soundness and completeness of the encoding in
Section~\ref{sec:sound_complete}.

Our encoding is based on the assumption that the resulting CHR programs are
executed only for initial states that correspond to graphs. We are not interested
in executions for arbitrary CHR states.

\subsubsection{States Encoding Graphs}
\label{sec:graph_states}

In this section we compare the different equivalence notions, i.e. graph
isomorphism and CHR state equivalence, and present a formal characterization of a
CHR state~$\sigma$ that is the encoding of a graph~$G$.

In order to determine if a CHR state encodes a graph, we define a predicate that
holds if and only if this is the case. It is intuitively clear, that starting
with the encoding of a graph and transforming it via a graph transformation rule
yields the encoding of a graph again. Formally, this is an invariant according to
the following definition. The first appearance of invariants in CHR research is
found in \cite{lam_sulz_linear_logic_agents_chr06} in the context of agent
programming.

\begin{definition}[Invariant]\label{def:invariant}

An \emph{invariant}~$\mcI$ is a predicate such that for all $\sigma_0$ and
$\sigma_1$, we have that if $\sigma_0 \der \sigma_1$ (or $\sigma_0 \equiv
\sigma_1$) and $\mcI(\sigma_0)$ then $\mcI(\sigma_1)$.
\end{definition}

The definition below introduces our desired property for CHR states. Note that it
is referred to as an invariant here, although we do not require it to be an
invariant throughout this section. In Section~\ref{sec:sound_complete}, more
precisely Corollary~\ref{cor:g_invariant}, we will show that it is indeed a
proper invariant.

\begin{definition}[Graph Invariant]\label{def:inv_graph}

Let $\sigma = \st{\bbG}{\bbB_c \land \bbB_a}{\bbV}$ be a state where $\bbB_c$ are
constraints of the form~$X=c$ for constants~$c$ and $\bbB_a$ are constraints of
the form~$X=Y{+}c_1{-}c_2$ for constants~$c_1,c_2$.

The \emph{graph invariant}~$\mcG$ holds for state $\sigma$ if and only if there
exists a graph~$G$ and a conjunction~$B$ of equality constraints of the
form~$X=c$ for a variable~$X$ and constant~$c$, such that \[ \st{\bbG}{\bbB_c
\land \bbB_a \land B}{\emptyset} \equiv \st{\chr(\gnd, G)}{\top}{\emptyset}\]

For a state~$\sigma$ for which $\mcG(\sigma)$ holds with a graph~$G$ we say
$\sigma$ is a $\mcG$-\emph{state~based~on}~$G$.
\end{definition}

\begin{example}

Consider again the final state~$\sigma''$ from the example computation in
Section~\ref{sec:encoding_example}:
\[
\sigma'' = \st{\node(N_3,D_3) \uplus \edge(E', N_3, N_3)}{\top}{\{N_3,D_3\}}
\]

By using the equality constraint~$B = (D_3 = 2)$ the resulting state for
Definition~\ref{def:inv_graph} is equivalent to: \[
\st{\node(N_3,2) \uplus \edge(E', N_3, N_3)}{\top}{\emptyset}
\]

Let $G$ be the graph consisting of a node~$v$ with a loop, then $\chr(\gnd, G) =
\node(N_v, 2) \uplus \edge(\tilde{E}, N_v, N_v)$. Therefore, the invariant~$\mcG$
is satisfied for the above state~$\sigma''$ as the corresponding states are
equivalent by renaming of local variables.

This example further shows why the variable set~$\bbV$ is disregarded for the two
states. The variable given by $\var$ for a node of the graph has to coincide with
the corresponding global variable for both states to be equivalent. Hence, for
the above graph with node~$v$, knowledge of the state~$\sigma''$ would be
necessary to determine that $\var(v) = N_3$. Omitting global variables from both
states, however, allows us to freely map~$v$ to any variable through $\var(v)$.
\end{example}

\begin{example} 

For the state~$\sigma = \st{\chr(\kp, G)}{\top}{\bbV}$ there clearly exists such
a graph~$G$, for which $B$ simply assigns the corresponding degree variables.
States may also be in-between $\chr(\gnd, G)$ and $\chr(\kp, G)$ in the sense
that only some of the degree variables are instantiated, resulting in a state~$\sigma'
= \st{\chr(\kp,G)}{\bbB_c}{\bbV}$ with $\bbB_c$ being the corresponding equality
constraints. By instantiating the remaining degrees it is clear that
$\mcG(\sigma')$ holds.
\end{example}

Note that arithmetic built-in constraints, introduced by bodies of rules in order
to adjust a node's degree, are covered by the above graph invariant definition:
The introduction of the corresponding degree equality constraint leads to a
collapse of the chain of arithmetic constraints. Hence, the concept of a
\mcG-state~based~on~$G$ also applies to intermediate computation states, which
gives rise to the following lemma.

\begin{lemma}[Graph States]\label{lem:graph_states} Let $\mcG(\sigma)$ hold for a
state~$\sigma$, then there exists a graph~$G$ such that \[ \sigma \equiv
\st{\chr(\kp, G)}{\bbBca}{\bbV} \]
\begin{itemize}
  \item $\bbB_c$ is a conjunction of $\dvar(v) = \deg_G(v)$ constraints
  \item $\bbB_a$ is a conjunction of $\dvar(v') = \dvar(v){+}c_1{-}c_2$
  constraints
\end{itemize}
\begin{proof}

Let $\sigma = \st{\bbG}{\bbBca}{\bbV}$, then by Def.~\ref{def:inv_graph} we have
that $\st{\bbG}{\bbBca \land B}{\emptyset} \equiv
\st{\chr(\gnd,G)}{\top}{\emptyset}$ for a graph~$G$ and $X=k$ constraints~$B$.

W.l.o.g. all identifier variables occurring in $\chr(\gnd,G)$ (and therefore in
$\chr(\kp,G)$) also occur in $\bbG$ as identifier variables. Due to the state
equivalence the difference between \bbG\ and $\chr(\kp,G)$ can then only consist
in \bbG\ specifying some node degrees by constants (for degree variables we can
again assume that they are the same as in $\chr(\kp,G)$).

Let $\Theta$ be a conjunction of equality constraints of the form~$X=c$ for each
degree specified explicitly in \bbG, using fresh variables for $X$. Interpreting
$\Theta$ as a substitution, replacing $X$ with $c$ for each of the equivalences,
we have that \[ \sigma \equiv \st{\chr(\kp,G) \Theta}{\bbBca}{\bbV}.\] As all
variables occurring in $\Theta$ are local, we get by
Def.~\ref{def:equiv}:\[\begin{array}{lcl}
                       \sigma & \eqct & \st{\chr(\kp,G) \Theta}{\bbBca \land
                       \Theta}{\bbV} \\
                        & \eqsubst & \st{\chr(\kp,G)}{\bbBca \land \Theta}{\bbV}\\
                        & = & \st{\chr(\kp,G)}{\bbB_c' \land \bbB_a}{\bbV}
                        \end{array}\]
\end{proof}
\end{lemma}

The reverse direction of Lemma~\ref{lem:graph_states} does not hold in
general: The state~$\sigma = \st{\emptyset}{D=0 \land X=1 \land
D=X{+}2{-}0}{\emptyset}$ satisfies the conditions for an empty graph~$G$, but of
course $\mcG(\sigma)$ does not hold, as $\st{\emptyset}{\bot}{\emptyset} \not
\equiv \st{\emptyset}{\top}{\emptyset}$.

The following lemma presents an interesting fact of the correspondence between
state equivalence and graph isomorphism: equivalent CHR states encoding two
graphs imply that these graphs are isomorphic.

\begin{lemma}[Equivalent \mcG-states imply Graph Isomorphism]
\label{lem:eq_iso}

Given a state $\sigma_1 = \st{\chr(\kp, G_1)}{\bbB_1}{\bbV}$, a
\mcG-state~based~on~$G_1$, and a state $\sigma_2 = \st{\chr(\kp,
G_2)}{\bbB_2}{\bbV}$, a \mcG-state~based~on~$G_2$, then \[ \sigma_1 \equiv
\sigma_2 \Rightarrow G_1 \simeq G_2 \]
\begin{proof}

First, we note that $\bbB_1,\bbB_2$ consist only of degree equalities or
adjustments. Therefore, we consider the following states instead, which are
already sufficient to imply the isomorphism: \[
\st{\chr(\kp, G_1)}{\top}{\bbV} \equiv \st{\chr(\kp, G_2)}{\top}{\bbV}
\] W.l.o.g. let the local variables occurring in the two states be disjoint (it
is clear that otherwise we can consider equivalent states that only differ by
renaming of local variables and that these states all provide corresponding graph
isomorphisms).

Let $\bar y_1$ and $\bar y_2$ be the set of local variables of the two states. We
can then apply the criterion from Thm.~\ref{thm:equiv_tf} to get \[ \CT \models
\exists \bar y_1 . \chr(\kp, G_1) = \chr(\kp, G_2). \] As there are only variable
terms contained in this equivalence we have the following conclusion, where
$c(\bar t)$ is any constraint with argument terms, i.e. variables,~$\bar t$. \[
\exists f : \bar y_1 \rightarrow \bar y_2 \text{ with } c(\bar t) \in \chr(\kp,
G_1) \rightarrow c(f(\bar t)) \in \chr(\kp, G_2) \] We know that $f$ is
surjective (as the variables are disjoint and the above equality demands that at
least one variable from $\bar y_1$ is mapped to each variable in $\bar y_2$). A
consequence of this is that $|\bar y_1| \ge |\bar y_2|$.

Analogously, we get from $\CT \models \exists y_2 . \chr(\kp, G_1) = \chr(\kp,
G_2)$ that $|\bar y_2| \ge |\bar y_1|$, and hence, $|\bar y_1| = |\bar y_2|$.
From this follows that $f$ is also injective, and therefore, bijective.

Next we realize, that by the above equality, $f$ has to map local variables
corresponding to node identifiers to local variables that also correspond to node
identifiers. Let $\bar y_{n1} \subset \bar y_1, \bar y_{n2} \subset \bar y_2$
be the local variables used for node identifiers, then $f' : \bar y_{n1}
\rightarrow \bar y_{n2}, y \mapsto f(y)$ is a well-defined and bijective
function. We use this to define the actual graph isomorphism function~$g :
V_{G_1} \rightarrow V_{G_2}$:
\[
g(v) = \left\{ \begin{array}{ll}
              v & \text{ if } \var(v) \in \bbV \\
              v' & \text{ if } \var(v) \in \bar y_{n1} \text{ and }
              f'(\var(v)) = \var(v')
              \end{array}\right.
\]
$g$ is well-defined: for every node there is a corresponding node identifier
variable and it has to be either global or local. If it is local, then $f'$ has
to map it to another local variable, as otherwise the $\equiv$ relation cannot
hold. Furthermore, $g$ is bijective as well, because it is defined
bijectively via $f'$ on local variables and the identity function on global
variables.

Finally, $g$ is a graph isomorphism: By the above equality we have corresponding
pairs of edge constraints. For every edge adjacent to a node given by a global
variable, the corresponding edge has to be adjacent to the same node with the
same global variable in order to satisfy $\equiv$. If the edge is adjacent to a
node identified by a local variable, then this variable is bijectively mapped to
another local variable and the above equality ensures that the corresponding edge
is adjacent to the same node as well. 
\end{proof}
\end{lemma}

The reverse direction of Lemma~\ref{lem:eq_iso} cannot hold in general:
The encoding of the graphs~$G_1$ and $G_2$ are independent from determining the
set~\bbV\ of global variables. Even a graph consisting of a single node only can
be encoded in two ways, such that the states are not equivalent:
\[
\st{\node(N, 0)}{\top}{\emptyset} \not \equiv \st{\node(N, 0)}{\top}{\{N\}}
\]

As indicated in Section~\ref{sec:encoding_example}, states may contain node
encodings with a variable degree. As these states are fundamental for program
analysis the following definition characterizes the set of these nodes.

\begin{definition}[Strong Nodes]\label{def:strong_state}

For a CHR state~$\sigma \equiv \st{\chr(\kp,G)}{\bbB}{\bbV}$ which is a
\mcG-state~based~on~$G$ we define the set of \emph{strong nodes} as: \[
\strong(\sigma) = \{ v \in V_G \mid \dvar(v) = \deg_G(v) \not \in \bbB\} \]
\end{definition}

The effect of strong nodes on computations and their use in program analysis is
discussed in detail in Section~\ref{sec:encoding_discussion}.

\subsubsection{Soundness and Completeness}
\label{sec:sound_complete}

In this section, we prove soundness and completeness of our embedding. That
$\mcG$ is an invariant for a GTS-CHR program and that termination of a GTS and
its GTS-CHR program coincide, are then derived as consequences of the main
theorem below.

\begin{theorem}[Soundness and Completeness]\label{thm:sound_and_complete}

Let $\sigma \equiv \st{\chr(\kp,G)}{\bbB}{\bbV}$ be a CHR state with
$\mcG(\sigma)$ holding with graph~$G$. Then \[ G \stackrel{r,m}{\Longrightarrow}
H \text{ with }  \{v \in V_G \mid \track_{G \dergts H}(v) \text{ defined}\}
\supseteq \strong(\sigma)\]
\centerline{if and only if}
\[ \sigma \der^r \tau \equiv \st{\chr(\kp,H)}{\bbB'}{\bbV} \text{ and }
\mcG(\tau) \text{ holds with graph } H.\]

\begin{proof}

In order to shorten this proof we use $k(G)$ and $g(G)$ to denote $\chr(\kp,G)$
and $\chr(\gnd,G)$, respectively.

``$\Longrightarrow$'':

Let $G \stackrel{r,m}{\Longrightarrow} H$ and let $r: L \leftarrow K
\rightarrow R$.

Let $\bbG := k(G) = k(G \setminus m(L)) \uplus k(m(E_L)) \uplus k(m(V_K)) \uplus
k(m(V_L \setminus V_K)) \Rightarrow \sigma \equiv \st{\bbG}{\bbB}{\bbV}$.

Let $\chrrule(r) = (r\ @\ C_L \Leftrightarrow C_R^b,C_R^u)$ with $C_L = k(K)
\uplus g(L\setminus K)$.

For $v \in V_L$ we have $\type_G(v)(\var(v),\_) \in C_L$ and\\
$\type_G(v)(\var(m(v)), \dvar(m(v))) \in k(m(V_K))$, as the types match due to
$m$ being a graph morphism.

As we have a fresh rule using node~$v$ that does not occur elsewhere we can say
that $\sigma \eqct \st{\bbG}{\var(m(v)) = \var(v) \land \bbB}{\bbV}$, and hence
\begin{eqnarray}
\label{eq:sigma_subst}
\sigma \eqsubst
\st{\bbG[\var(m(v))/\var(v)]}{\var(m(v)) = \var(v) \land \bbB}{\bbV}
\end{eqnarray}

Consider $v \in V_L \setminus V_K$, then $\type_G(v)(\var(v),\deg_L(v)) \in
C_L$. Assume that $m(v) \in \strong(\sigma)$, then $\track_{G \dergts H}(m(v))$
is defined, which is a contradiction to $v \in V_L \setminus V_K$. Therefore, $m(v)
\not \in \strong(\sigma)$ and hence $\dvar(m(v)) = \deg_G(m(v)) \in \bbB$. As
$G \stackrel{r,m}{\Longrightarrow} H$ satisfies the gluing condition, we know 
that $\deg_L(v) = \deg_G(m(v))$. Therefore, we have that
\begin{eqnarray*}
\sigma\eqsubst
\st{\bbG[\var(m(v))/\var(v)][\dvar(m(v))/\deg_G(m(v))]}{\\\var(m(v))=\var(v)
\land \bbB}{\bbV}
\end{eqnarray*}

From \eqref{eq:sigma_subst} for nodes~$v \in V_K$ and the above for nodes~$v \in
V_L \setminus V_K$ follows for a conjunction of equality constraints~$E$ that \[
\sigma \equiv \st{k(G\setminus m(L)) \uplus k(m(E_L)) \uplus k(V_K) \uplus g(V_L
\setminus V_K)}{\bbB \land E}{\bbV} =
\st{\bbG'}{\bbB \land E}{\bbV}
\]

Let $e \in E_L$, than $\type_G(e)(\var(e),\var(\src(e)),\var(\tgt(e))) \in C_L$
and after the previous substitutions have been made for node identifier
variables, and as $k(e) = g(e)$, we get
$\type_G(m(e))(\var(m(e)),\var(\src(e)),\var(\tgt(e))) \in \sigma$ . We then have
\begin{eqnarray}\label{eq:sigma_subst3}
\sigma \equiv
\st{\bbG'[\var(m(e))/\var(e)]}{\var(m(e)) = \var(e) \land \bbB \land E}{\bbV}
\end{eqnarray}

By applying this substitution for all edges~$e \in E_L$ and extending $E$ with
the required equalities to $E'$ we get: \[ \sigma \equiv \st{k(G\setminus m(L))
\uplus k(E_K) \uplus g(E_L \setminus E_K) \uplus k(V_K) \uplus g(V_L \setminus
V_K)}{\bbB \land E'}{\bbV} \] Hence, $\sigma \equiv \st{k(G \setminus m(L))
\uplus C_L}{\bbB \land E'}{\bbV}$ such that we apply the rule~$\chrrule(r)$ to
$\sigma$: \[
\begin{array}{lcl}
 \sigma \der^r \tau & \equiv & \st{k(G \setminus m(L))
\uplus C_R^u}{\bbB \land E' \land C_R^b}{\bbV}\\
& \equiv & \st{k(G \setminus m(L)) \uplus g(R \setminus K) \uplus k(E_K) \uplus
k(V_K')}{\bbB \land E' \land C_R^b}{\bbV}
\end{array}
\]

As $C_R^b$ contains $\var(v') = \var(v) \forall v \in V_K$ let $C_R'$ be
$C_R^b$ without these constraints, then
\[
\begin{array}{lcl}
\tau & \eqsubst & \st{k(G \setminus m(L)) \uplus g(R \setminus K) \uplus k(E_K)
\uplus k(V_K)}{\bbB \land E' \land C_R^b}{\bbV} \\
 & \eqct & \st{k(G \setminus m(L)) \uplus g(R \setminus K) \uplus k(E_K) \uplus
 k(V_K)}{\bbB \land E' \land C_R'}{\bbV}
\end{array}
\]

Let $\hat{\bbG} := k(G\setminus m(L)) \uplus k(K)$, then $\tau \eqct
\st{\hat{\bbG} \uplus g(R \setminus K)}{\bbB \land E' \land C_R' \land
D_R}{\bbV}$ with $\forall v \in V_R \setminus V_K . \dvar(v) = \deg_R(v) \in
D_R$. Furthermore, consider $\Theta$ a substitution corresponding to the reverse
reading of $E'$ which undoes the ideas of \eqref{eq:sigma_subst} and
\eqref{eq:sigma_subst3} for all affected nodes and edges. We then get

\[
\begin{array}{lcl}
\tau & \eqsubst & \st{\hat{\bbG} \uplus k(R \setminus K)}{D_R \land C_R' \land
\bbB \land E'}{\bbV} \\
& \eqsubst & \st{k(G \setminus m(L \setminus K)) \uplus
(k(R\setminus K) \Theta)}{D_R \land C_R' \land \bbB \land E'}{\bbV}\\
& \eqct & \st{k(G \setminus m(L \setminus K)) \uplus
(k(R\setminus K) \Theta)}{D_R \land C_R'\Theta \land \bbB}{\bbV}\\
& \equiv & \st{k(H)}{\bbB'}{\bbV}
\end{array}\]

We get the graph~$H$ as its DPO construction corresponds to the removal of $m(L
\setminus K)$ and addition of $R\setminus K$. $\Theta$ is needed to attach the
new nodes of $R \setminus K$ to nodes from $V_K$ and $C_R'$ contains degree
adjustments for those nodes that are correct by construction. Hence, it also
holds that $\mcG(\tau)$ is satisfied with graph~$H$.

\vskip1cm
``$\Longleftarrow$'':

Let $\sigma \der^r \tau$ with $\tau \equiv \st{k(H)}{\bbB'}{\bbV}$ and
$\mcG(\tau)$ holds with graph~$H$. Let $\chrrule(r) = (r\ @\ C_L \Leftrightarrow
C_R^b,C_R^u$ with $C_L = k(K) \uplus g(L \setminus K)$. From Def.~\ref{def:opsem}
follows that
\begin{eqnarray} \label{eq:sigma_rule}
\sigma \equiv \st{k(K) \uplus g(L
\setminus K) \uplus k(G \setminus L)}{\bbB_1}{\bbV}
\end{eqnarray}

Using Lemma~\ref{lem:graph_states} and with $E$ being a conjunction of
$\var(m(x)) = \var(x)$ constraints for $x \in L$ we get: \[
\begin{array}{lcl}
\sigma & \equiv & \st{k(G)}{\bbBca}{\bbV} \\
& \equiv & \st{k(K) \uplus k(L \setminus K) \uplus k(G\setminus m(L))}{\bbBca
\land E}{\bbV} \\
& \stackrel{\eqref{eq:sigma_rule}}{\equiv} & \st{k(K) \uplus g(L \setminus K)
\uplus k(G \setminus m(L))}{\bbBca \land E'}{\bbV}
\end{array}\]
where $E'$ is the extension of $E$ with $\dvar(m(v)) = \deg_G(v)$ constraints
for $v \in V_L \setminus V_K$ and $\bbB_1 = \bbBca \land E'$.

$m: L \rightarrow G$ is well-defined and injective by the multiset semantics of
CHR and it remains to be shown, that $m$ is a graph morphism. Therefore, let $e
\in E_L$, then $\type_L(e)(\var(e), \var(\src(e)), \var(\tgt(e))) \in C_L$ and
$\type_L(\src(e))(\var(\src(e)), \_) \uplus \type_L(\tgt(e))(\var(\tgt(e)), \_)
\in C_L$. Hence, $\var(m(e)) = \var(e)$, $\var(m(\src(e))) = \var(\src(e))$ and
$\var(m(\tgt(e))) = \var(\tgt(e))$ are all in $\bbB_1$. Therefore, $m(\src(e)) =
\src(m(e)) \land m(\tgt(e)) = \tgt(m(e))$.

The gluing condition is satisfied, as $\forall v \in V_L \setminus V_K$ the
matched degree ensures that there are no dangling edges, hence, $r$ is
GTS-applicable to $G$. Similarly to the other proof direction, we show that the
DPO construction of $H$ coincides with the construction of $\tau$ by CHR rule
application:

\[\begin{array}{lcl} \sigma \der^r \tau & \equiv & \st{k(K) \uplus g(R \setminus
K) \uplus k(G
\setminus m(L))}{\bbBca \land E' \land C_R^b}{\bbV}\\
& \eqct & \st{k(K) \uplus g(R \setminus K) \uplus k(G
\setminus m(L))}{\bbBca \land E \land C_R^b}{\bbV}\\
& \eqsubst & \st{k(m(K)) \uplus g(R \setminus K) \uplus k(G
\setminus m(L))}{\bbBca \land E \land C_R^b}{\bbV}\\
& = & \st{g(R \setminus K) \uplus k(G \setminus m(L \setminus K))}{\bbBca \land
E \land C_R^b}{\bbV} \\
& \equiv & \st{g(R \setminus K) \Theta \uplus k(G \setminus m(L \setminus
K))}{\bbBca \land C_R^b}{\bbV}\\
& \equiv & \st{k(H)}{\bbB'}{\bbV}
  \end{array}
\] where $\Theta$ is the reverse substitution for $E$ similar to the other proof
direction. The final equivalence comes from extracting the degrees of constraints
in $g(R \setminus K)$ into equality constraints contained in $\bbB'$. As can be
seen here, the application of the rule results in a state encoding the graph~$H$,
such that $\mcG(\tau)$ holds.

Finally, for the set~$\strong(\sigma)$ we know that the nodes cannot be removed
by rule~$r$: For a node~$v\in V_L\setminus V_K$ we have
$\type_L(v)(\var(v),\deg_L(v)) \in C_L$, but this cannot be matched with
$\sigma$, as by Def.~\ref{def:strong_state} the corresponding degree is
unavailable. Hence, none of the nodes from $\strong(\sigma)$ are removed by the
rule application~$G \stackrel{r,m}{\Longrightarrow} H$, i.e. $\track_{G
\dergts H}(v)$ is defined for all $v \in \strong(\sigma)$.
\end{proof}
\end{theorem}

As can be seen in the proof of Theorem~\ref{thm:sound_and_complete}, a GTS-CHR
rule application on a \mcG-state~based~on~$G$ always results in a state encoding
a corresponding graph~$H$, which gives us the following corollary.

\begin{corollary}[\mcG\ Invariant]\label{cor:g_invariant}

For a GTS-CHR program \mcG\ is an invariant.
\end{corollary}

A closer look at the conditions required in Theorem~\ref{thm:sound_and_complete}
reveals that for a state~$\sigma$ with $\strong(\sigma) = \emptyset$, i.e. for an
encoding of a graph with all degrees explicitly given, we have unrestricted
soundness and completeness.

\begin{corollary}[Unrestricted Soundness and Completeness]
\label{cor:sound_and_complete}

Let $\sigma \equiv \st{\chr(\gnd,G)}{\top}{\bbV}$ be a CHR state with
$\mcG(\sigma)$ holding with graph~$G$. Then \[ G \stackrel{r,m}{\Longrightarrow}
H\]
\centerline{if and only if}
\[ \sigma \der^r \tau \equiv \st{\chr(\gnd,H)}{\top}{\bbV} \text{ and }
\mcG(\tau) \text{ holds with graph } H\]
\begin{proof}
This follows from Theorem~\ref{thm:sound_and_complete} and the following
insight: as all degrees of $G$ are specified explicitly and all nodes added by
the rule are also given explicit degrees, all degrees in $H$ are given
explicitly as well, which allows us to use $\chr(\gnd,H)$ here.
\end{proof}
\end{corollary}

Finally, the soundness and completeness result induces a termination
correspondence between a GTS and its GTS-CHR program. Again, we restrict our
observation to graph-encoding states.

\begin{corollary}[Termination Correspondence]\label{cor:termination} A GTS is
terminating if and only if its corresponding GTS-CHR program is \mcG-terminating,
i.e. terminating for all \mcG-states.
\begin{proof}
If a GTS contains a non-terminating derivation, we have the corresponding
computation in its GTS-CHR program by Corollary~\ref{cor:sound_and_complete}.
Similarly, if the GTS-CHR program has a non-terminating computation, there
exists a corresponding non-terminating GTS derivation according to
Theorem~\ref{thm:sound_and_complete}.
\end{proof}
\end{corollary}

\subsection{Discussion}
\label{sec:encoding_discussion}

In this section we discuss our previously presented encoding. First,
Section~\ref{sec:partial} investigates that a GTS-CHR program works with
partially defined graphs and explains the suitability of these graphs for program
analysis. Then we present ways to simplify the encoding of GTS-CHR rules in
Section~\ref{sec:diff_encoding}.

\subsubsection{Partially Defined Graphs}
\label{sec:partial}

In the example computation given in Section~\ref{sec:encoding_example} the input
contains a node with a variable degree: $\node(N_3, D_3)$. Nevertheless,
computations on this input are possible and the example resulted in the final
state: \[\st{\node(N_3, D_3) \uplus \edge(E', N_3, N_3)}{\top}{\{N_3,
D_3\}}\]

In general, a variable node degree will cause a chain of degree adjustment
constraints to be created, i.e. constraints of the form $X = Y {+} c_1 {-} c_2$.
These stem from the node being involved in a rule application that affects its
degree.

It is important to realize that we can only match such a node in rules that do
not remove it. A rule that removes a node contains the explicit degree for that
node in the head, which cannot be matched through a variable degree. As a
consequence, specifying variable degrees in the input ensures that the
corresponding nodes will not be removed by the computation. This also becomes
clear from the investigation of strong nodes in the previous section.

While this is an interesting feature in its own right, it provides the basis for
many forms of program analysis. The aim of program analysis is to make statements
on an infinite number of graphs, while only having to investigate a small
selection of graphs. Graph encodings with variable degrees can here be thought of
as partially defined graphs, i.e. there may be any number of further edges being
connected to a node with a variable degree. 

Note that partially defined graphs only exist within the CHR context. In a GTS
the degree of a node is implicitly given by the adjacent edges. As a consequence,
leaving a node's degree undefined in the CHR encoding ensures, that this node
will not be removed during computation. In the GTS context we have no such option
available for host graphs.

By the above argument, the state \[ \st{\node(N, D)}{\top}{\{N,D\}}\] therefore
not only represents the graph consisting of a single node and no edges. Instead,
it represents the set of all graphs with at least one node. Similarly, the above
final state from Section~\ref{sec:encoding_example} stands for the set of graphs
that contain at least one node with a loop.

Every computation performed on an input with variable degrees actually represents
computations for an infinite set of graphs. This is a fundamental feature for the
usage of our encoding in program analysis and will be exploited in
Sections~\ref{sec:confluence} and \ref{sec:opeq}.

\subsubsection{Different Encoding Possibilities}
\label{sec:diff_encoding}

The encoding proposed in this work can be varied in several different ways. We
chose the encoding in Definition~\ref{def:encoding} and
Definition~\ref{def:chr_rule} for this work, because it is a verbose encoding,
hence, directly presenting all its components and simplifying the proofs. In
practice however, a less verbose encoding resulting in shorter rules can be used
instead. In this section we present different possible simplifications achieving
this.

The different simplifications are illustrated by applying them to the
\emph{twoloop} rule which is of the following form when encoded as specified in
Definition~\ref{def:chr_rule}:

\begin{center} 
\begin{tabular}{ll} 
twoloop $@$ & $\node(N_1, D_1) \uplus \node(N_2, 2) \uplus$\\
& $\edge(E_1, N_1, N_2) \uplus \edge(E_2, N_2, N_1)$\\
& $\Leftrightarrow$\\
& $\node(N_1', D_1') \uplus \edge(E_3, N_1, N_1), N_1' = N_1 \land D_1' =
D_1{-}2{+}2$
\end{tabular}
\end{center}

There are two ways to specify the degree of nodes in $L \setminus K$. The one
chosen in Definition~\ref{def:chr_rule} explicitly specifies the respective
degree in the head. Another way is to keep the degree as a variable~$D$ in the
head and add the built-in constraint $D=k$ to the guard of the rule. However,
most current CHR compilers detect these equalities and automatically transform
between them to the representation most suitable for an optimization. Therefore,
in this work we directly specify the degree in the head to avoid guards
altogether.

\paragraph{Variable Elimination}

As Definition~\ref{def:chr_rule} encodes a node~$v \in V_K$ using a new node
identifier~$v'$ with $\var(v) = \var(v')$ and $\var(v')$ is not used elsewhere,
this substitution can be included directly into the rule encoding:

\begin{center} 
\begin{tabular}{ll} 
twoloop $@$ & $\node(N_1, D_1) \uplus \node(N_2, 2) \uplus$\\
& $\edge(E_1, N_1, N_2) \uplus \edge(E_2, N_2, N_1)$\\
& $\Leftrightarrow$\\
& $\node(N_1, D_1') \uplus \edge(E_3, N_1, N_1), D_1' = D_1{-}2{+}2$
\end{tabular}
\end{center}

Note that we perform variable elimination on node identifiers by default in the
remainder of this work. However, as we need to take degree adjustments into
account, the formulation of Definition~\ref{def:chr_rule} is simplified by the
variable duplication.

\paragraph{Arithmetic Simplification}

The degree adjustments in Definition~\ref{def:chr_rule} explicitly contain the
information on how many edges the rule deletes and creates. For the adjustment
itself, however, it is sufficient to simply adjust the degree by the actual
change in the number of edges. Additionally, if the change is $0$, like in the
\emph{twoloop} rule, the extra local variable used for the degree can be
substituted, resulting in:

\begin{center} 
\begin{tabular}{ll} 
twoloop $@$ & $\node(N_1, D_1) \uplus \node(N_2, 2) \uplus$\\
& $\edge(E_1, N_1, N_2) \uplus \edge(E_2, N_2, N_1)$\\
& $\Leftrightarrow$\\
& $\node(N_1, D_1) \uplus \edge(E_3, N_1, N_1)$
\end{tabular}
\end{center}

\paragraph{Elimination of Edge Identifiers}

The edge identifier variables are used throughout this work, because they
simplify dealing with the multiset semantics of CHR with respect to the edge
constraint representing exactly one edge of a graph. In a CHR implementation,
however, every constraint is implemented as a unique object -- sometimes even
annotated with an identifier number -- which makes the explicit edge identifiers
redundant. Using this idea the \emph{twoloop} rule can be further simplified to:

\begin{center} 
\begin{tabular}{ll} 
twoloop $@$ & $\node(N_1, D_1) \uplus \node(N_2, 2) \uplus$\\
& $\edge(N_1, N_2) \uplus \edge(N_2, N_1)$\\
& $\Leftrightarrow$\\
& $\node(N_1, D_1) \uplus \edge(N_1, N_1)$
\end{tabular}
\end{center}

Note that the same argumentation cannot be applied to node identifiers, as those
are required for specifying the source and target of edge constraints.

\paragraph{Simpagation Rules}

Some nodes and edges of the left-hand rule graph~$L$ of a GTS rule can occur only
to specify a certain graph context and are unaffected by the rule application.
For nodes this can also happen if the modification to adjacent edges results in
no change to the degree, as in the \emph{twoloop} rule. In those cases, the node
or edge is encoded in exactly the same way in the head and body of the rule.
Therefore, during the rule application the corresponding constraint is removed
and introduced again. Using a simpagation rule allows us to move such a
constraint into the part of the head which is not removed during the rule
application. This reduces the textual size of the rule as well as its execution
time, because it avoids the generation of a new constraint during the rule
application.

After applying all the previous simplifications to the \emph{twoloop} rule and
transforming it into a simpagation rule we get the following simplified rule:

\begin{center} 
\begin{tabular}{ll} 
twoloop $@$ & $\node(N_1, D_1) \setminus$\\
& $\node(N_2, 2) \uplus \edge(N_1, N_2) \uplus \edge(N_2, N_1)$\\
& $\Leftrightarrow$\\
& $\edge(N_1, N_1)$
\end{tabular}
\end{center}

One might be tempted to always create simpagation rules in
Definition~\ref{def:chr_rule}, based on the idea that the context graph~$K$
already identifies non-removed nodes. However, the above creation of simpagation
rules with node constraints among the kept constraints, is only possible if the
respective node's degree remains unchanged by the rule application.

Readers more familiar with CHR may also wonder if propagation rules could be used
as well. It is technically possible to define a GTS rule that does not remove any
elements, but only adds new nodes and edges. However, a thusly created GTS would
suffer from a problem that in CHR literatue is referred to as trivial
non-termination (see e.g., \cite{fruehwirth09}), i.e. such a rule could be
applied infinitely often. For this reason, most CHR implementations restrict
propagation rule applications, hence, our encoding using simplification or
simpagation rules remains more faithful to the semantics of graph
transformations.

\section{Analyzing Confluence}
\label{sec:confluence}

The confluence property is relevant to both, graph transformation systems and
Constraint Handling Rules. It guarantees that any terminating computation made
for an initial state results in the same final state no matter in which order
applicable rules are applied.

In Section~\ref{sec:confl_gts} we formally introduce confluence, both for GTS and
CHR. Furthermore, we give the definitions for \emph{critical pairs} in both
systems, which are derived directly from the rules. Investigation of critical
pairs for determining confluence of a terminating rewrite system goes back to
research about term rewriting systems \cite{huet80}, and both, GTS and CHR, have
adapted the corresponding criteria.

Next, Section~\ref{sec:confl_pairs} examines the relation between critical pairs
of a GTS and its corresponding GTS-CHR program. We then introduce the concept of
\emph{observable confluence} \cite{duckstuckeysulzmann07}. It is a technical
means to restrict our observations to CHR states that correspond to graphs. This
in turn results in a closer correspondence between GTS and CHR for later results.

For terminating GTS, confluence analysis proved to be undecidable: \cite{plump05}
showed that the critical pair analysis gives only a sufficient criterion for
confluence. We show that the decidable observable confluence test of a GTS-CHR
program coincides with this criterion.

The discrepance in decidability of the two systems' confluence properties is
discussed in Section~\ref{sec:confl_discussion} for exemplary critical pair
analyses.

\subsection{Preliminaries}
\label{sec:confl_gts}

This subsection introduces the necessary definitions for GTS and CHR confluence
before comparing the two notions. Unless noted otherwise, the involved graph
transformation systems and GTS-CHR programs are assumed to be terminating.

\begin{definition}[GTS Confluence]A GTS is called \emph{confluent} if, for all
typed graph transformations $G \stackrel{*}{\Longrightarrow} H_1$ and $G
\stackrel{*}{\Longrightarrow} H_2$, there is a typed graph $X$ together with
typed graph transformations $H_1 \stackrel{*}{\Longrightarrow} X$ and $H_2
\stackrel{*}{\Longrightarrow} X$. \emph{Local confluence} means that this
property holds for all pairs of direct typed graph transformations $G \dergts
H_1$ and $G \dergts H_2$ \cite{ehrigprangetaentzer06}.
\end{definition}

Newman's general result for rewriting systems \cite{newman} implies that it is
sufficient to consider local confluence for terminating graph transformation
systems. To verify local confluence, we particularly need to study critical pairs
and their joinability, according to the following definition based on
\cite{ehrigprangetaentzer06,plump05}.

\begin{definition}[Joinability of Critical GTS Pair]\label{def:gts_cp} Let $r_1 =
(L_1 \stackrel{l}{\leftarrow} K_1 \stackrel{r}{\rightarrow} R_1), r_2 = (L_2
\stackrel{l}{\leftarrow} K_2 \stackrel{r}{\rightarrow} R_2)$ be two GTS rules. A
pair $P_1 \stackrel{r_1, m_1}{\Longleftarrow} G \stackrel{r_2,
m_2}{\Longrightarrow} P_2$ of direct typed graph transformations is called a
\emph{critical GTS pair} if it is parallel dependent, and minimal in the sense
that the pair $(m_1, m_2)$ of matches $m_1: L_1 \rightarrow G$ and $m_2 : L_2
\rightarrow G$ is jointly surjective.

A pair $P_1 \stackrel{r_1, m_1}{\Longleftarrow} G \stackrel{r_2,
m_2}{\Longrightarrow} P_2$ of direct typed graph transformations is called
\emph{parallel independent} if $m_1(L_1) \cap m_2(L_2) \subseteq m_1(K_1) \cap
m_2(K_2)$, otherwise it is called \emph{parallel dependent}.

A critical GTS pair $P_1 \stackrel{r_1, m_1}{\Longleftarrow} G \stackrel{r_2,
m_2}{\Longrightarrow} P_2$ is called \emph{joinable} if there exist typed graphs
$X_1,X_2$ together with typed graph transformations $P_1
\stackrel{*}{\Longrightarrow} X_1 \simeq X_2 \stackrel{*}{\Longleftarrow} P_2$.
It is \emph{strongly joinable} if there is an isomorphism~$f: X_1 \rightarrow
X_2$ such that for each node~$v$, for which $\track_{G \dergts P_1}(v)$ and
$\track_{G \dergts P_2}(v)$ are defined, the following holds:
\begin{enumerate}
  \item $\track_{G \dergts P_1 \dergts X_1}(v)$ and $\track_{G
  \dergts P_2 \dergts X_2}(v)$ are defined and
  \item $f_V(\track_{G \dergts P_1 \dergts X_1}(v)) = \track_{G
  \dergts P_2 \dergts X_2}(v)$
\end{enumerate}
\end{definition}

A similar notion of confluence has been developed for CHR. The following
definition is an adaptation of \cite{fruehwirth09} to the operational semantics
on equivalence classes.

\begin{definition}[CHR Confluence] A CHR program is called \emph{confluent} if
for all states $\sigma, \sigma_1,$ and $\sigma_2$: If $\sigma_1 \derrev \sigma \der^*
\sigma_2$, then $\sigma_1$ and $\sigma_2$ are joinable. Two states $\sigma_1$ and
$\sigma_2$ are called \emph{joinable} if there exists a state $\tau$ such that
$\sigma_1 \der^* \tau \derrev \sigma_2$.
\end{definition}

Analogous to a GTS, the confluence property for terminating CHR programs is
determined by local confluence which can be checked through critical pairs. The
following definition is adapted to the situation in this work, i.e. it only
considers simplification rules and no guards.

\begin{definition}[Joinability of Critical CHR Pair]\label{def:chr_cp} Let $r_i,
i =1,2$ be two (not necessarily different) simplification rules of the following
kind with variables that have been renamed apart: \[ H_i \uplus A_i
\Leftrightarrow B_i^u,B_i^b \]

Then an \emph{overlap}~$\sigcp$ of $r_1$ and $r_2$ is $\sigcp = \st{H_1 \uplus
A_1 \uplus H_2}{A_1 = A_2}{\bbV}$, provided $A_1$ and $A_2$ are non-empty
multisets, $\bbV = \vars(H_1 \uplus A_1 \uplus H_2 \uplus A_2)$ and $\CT \models
\exists (A_1 = A_2)$.

Let $\sigma_1 = \st{B_1^u \uplus H_2}{B_1^b \land (A_1 = A_2)}{\bbV}$ and
$\sigma_2 = \st{B_2^u \uplus H_1}{B_2^b \land (A_1 = A_2)}{\bbV}$. Then the
tuple~$\mcCP = (\sigma_1, \sigma_2)$ is a \emph{critical CHR pair} of $r_1$ and
$r_2$. A critical CHR pair $(\sigma_1, \sigma_2)$ is \emph{joinable} if
$\sigma_1$ and $\sigma_2$ are joinable.
\end{definition}

\subsection{Analyzing Confluence via Critical Pairs}
\label{sec:confl_pairs}

After defining the different notions of confluence we now further investigate the
difference between critical GTS pairs and critical CHR pairs for GTS-CHR
programs. The following lemma shows that there exists a corresponding overlap for
each critical GTS pair. Therefore, by examining the overlaps and using the
previous soundness result we can transfer joinability results to the critical GTS
pair.

\begin{lemma}[Overlap for Critical GTS Pair]\label{lem:cp_gts_chr}

If $P_1 \stackrel{r_1,m_1}{\Longleftarrow} G \stackrel{r_2,m_2}{\Longrightarrow}
P_2$ is a critical GTS pair, then there exists an overlap~$\sigma_{\mcCP}$
of $\chrrule(r_1) = (r_1\ @\ C_{L1} \Leftrightarrow C_{R1}^u,C_{R1}^b)$ and
$\chrrule(r_2) = (r_2\ @\ C_{L2} \Leftrightarrow C_{R2}^u,C_{R2}^b)$ which is a
\mcG-state~based~on~$G$ and a critical CHR pair~$(\sigma_1, \sigma_2)$ such that
$\sigma_1$ is a \mcG-state~based~on~$P_1$ and $\sigma_2$ is a
\mcG-state~based~on~$P_2$.

\begin{proof}

Let the two GTS rules be $L_i \leftarrow K_i \rightarrow R_i$ for $i=1,2$ and let
$M = m_1(L_1) \cap m_2(L_2)$. We then define the following sets of constraints
from which we construct the overlap: \[
\begin{array}{lcl}
H_1 & = & \{\chr_{L_1}(\kp, x) \mid x \in L_1 \land m_1(x) \not \in M \}\\
H_2 & = & \{\chr_{L_2}(\kp, x) \mid x \in L_2 \land m_2(x) \not \in M \} \\
A_1 & = & \{\chr_{L_1}(\kp, x) \mid x \in L_1 \land m_1(x) \in M \} \\
A_2 & = & \{\chr_{L_2}(\kp, x) \mid x \in L_2 \land m_2(x) \in M \}\\
C_1 & = & \{\dvar(v) = \deg_{L_1}(v) \mid v \in V_{L_1} \setminus V_{K_1}\}\\
C_2 & = & \{\dvar(v) = \deg_{L_2}(v) \mid v \in V_{L_2} \setminus V_{K_2}\}
\end{array}
\]

Let $\bbV = \vars(H_1 \uplus H_2 \uplus A_1 \uplus A_2)$ and let $\sigma =
\st{H_1}{C_1}{\bbV}$, then $\sigma \equiv \sigma' = \st{\{\chr_{L_1}(\kp,x) \mid
x \in K_1 \land m_1(x) \not \in M\} \uplus \{\chr_{L_1}(\gnd,x) \mid x \in L_1
\setminus K_1 \land m_1(x) \not \in M\}}{\top}{\bbV} =: \st{H_2'}{\top}{\bbV}$ by
applying $C_1$ as a substitution to $H_1$, and then removing $C_1$ as all
$\dvar(v)$ variables for $v \in V_{L_1} \setminus V_{K_1}$ are then strictly
local.

Similarly, $\st{A_1}{C_1}{\bbV} \equiv \st{\{\chr(\kp,x) \mid x \in K_1 \land
m_1(x) \in M\} \uplus \{\chr(\gnd,x) \mid x \in L_1 \setminus K_1 \land m_1(x)
\in M}{\top}{\bbV} =: \st{A_1'}{\top}{\bbV}$, and analogously, we define $H_2'$
and $A_2'$.

By Def.~\ref{def:chr_rule} we have that $H_1' \uplus A_1' = C_{L1}$  and $H_2'
\uplus A_2' = C_{L2}$. As $M \ne \emptyset$ it follows that $A_1'$ and $A_2'$ are
non-empty. To investigate if $\CT \models \exists (A_1' = A_2')$ we take a closer
look at the equality constraints imposed by $A_1' = A_2'$: \[
\begin{array}{cl}
& \{\var(v_1) = \var(v_2) \mid v_1 \in V_{L_1} \land v_2 \in
V_{L_2}, m_1(v_1) = m_2(v_2) \} \\
\land & \{\dvar(v_1) = \dvar(v_2) \mid v_1 \in V_{K_1} \land v_2 \in V_{K_2}
\land m_1(v_1) = m_2(v_2) \}\\
\land & \{\dvar(v_1) = \deg_{L_2}(v_2) \mid v_1 \in V_{K_1} \land v_2 \in
V_{L_2} \setminus V_{K_2} \land m_1(v_1) = m_2(v_2) \} \\
\land & \{\dvar(v_2) = \deg_{L_1}(v_1) \mid v_1 \in V_{L_1} \setminus
V_{K_1} \land v_2 \in V_{K_2} \land m_1(v_1) = m_2(v_2) \} \\
\land & \{\var(e_1) = \var(e_2) \mid e_1 \in E_{K_1} \land e_2 \in E_{K_2}
\land m_1(e_1) = m_2(e_2) \}\\
\land & \{\deg_{L_1}(v_1) = \deg_{L_2}(v_2) \mid v_1 \in V_{L_1} \setminus
V_{K_1} \land v_2 \in V_{L_1} \setminus V_{K_2} \land\\& m_1(v_1) = m_2(v_2) \}
\\
\end{array}
\]

Except for the last row, the above equality constraints can easily be satisfied
under existential quantification. Hence, the only remaining problematic case is
when two node constraints with constant degrees are overlapped. However, the
degree of $m_1(v_1) = m_2(v_2)$ equals the degree of $v_1$ and the degree of
$v_2$ due to the gluing condition being satisfied, such that this case can only
occur with equal constant degrees.

Hence, $\sigma_{\mcCP} = \st{H_1' \uplus A_1' \uplus H_2'}{A_1' = A_2'}{\bbV}$ is
an overlap of $\chrrule(r_1)$ and $\chrrule(r_2)$ with the critical CHR
pair~$(\st{C_{R1}^u \uplus H_2'}{A_1' = A_2' \land C_{R1}^b}{\bbV},
\st{C_{R2} \uplus H_1'}{A_1' = A_2' \land C_{R2}^b}{\bbV}$.\end{proof}

\end{lemma}

If we try to directly transfer the confluence property of a GTS to the
corresponding GTS-CHR program, we cannot succeed however, as in general there are
too many critical CHR pairs that could cause the GTS-CHR program to become
non-confluent. The following example provides a rule which only has one critical
GTS pair, but for which the corresponding CHR rule has three critical CHR pairs.

\begin{figure}
\includegraphics{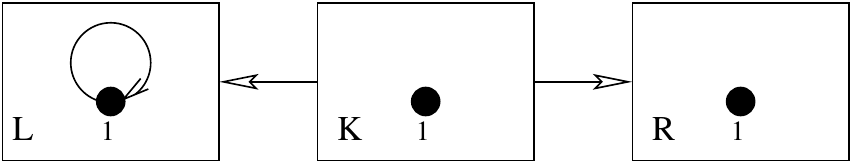}
\caption{Graph production rule for removing a loop} 
\label{fig:remloop}
\end{figure}

\begin{example}\label{ex:confluence}
Consider the graph production rule in Fig.~\ref{fig:remloop}. It removes a
loop from a node and has the following corresponding CHR rule:\[
R\ @\ \node(N, D) \uplus \edge(E, N, N) \Leftrightarrow \node(N, D'), D'
= D-2\]

To investigate confluence one must overlap this rule with itself which yields the
following three CHR overlap states:
\begin{enumerate}
  \item \label{ex:conf:s1} $\st{\node(N,D) \uplus \edge(E,N,N) \uplus
  \edge(E',N',N')}{N = N'}{\{N,D,E,E',N'\}}$
  \item \label{ex:conf:s2} $\st{\node(N,D) \uplus \node(N',D') \uplus
  \edge(E,N,N)}{N=N'}{\{N,D,N',D',E\}}$
  \item \label{ex:conf:s3} $\st{\node(N,D) \uplus
  \edge(E,N,N)}{\top}{\{N,D,E\}}$
\end{enumerate}

State~\eqref{ex:conf:s1} is not critical, because the corresponding pair of graph
transformations is parallel independent (cf. \cite{ehrigprangetaentzer06}, and
hence, directly joinable by applying the rule again. State~\eqref{ex:conf:s2} is
an invalid state, i.e. it violates \mcG, as it has multiple encodings of the same
node and state~\eqref{ex:conf:s3} is the encoding of the corresponding critical
pair for the graph production rule.
\end{example}

As we want to rule out invalid states, we use the following notion of observable
confluence presented in \cite{duckstuckeysulzmann07}. It is based on restricting
confluence investigations to states that satisfy an invariant. Based on these
invariants, observable confluence (or \mcI-confluence) is defined as follows:

\begin{definition}[Observable Confluence]\label{def:obs_conf}

A CHR program $P$ is \mcI-confluent with respect to invariant~\mcI\ if the
following holds for all states $\sigma, \sigma_1$, and $\sigma_2$ where
$\mcI(\sigma)$ holds: If $\sigma_1 \derrev \sigma \der^* \sigma_2$ then
$\sigma_1$ and $\sigma_2$ are joinable.
\end{definition}

In order to use the graph invariant~\mcG\ for the notion of observable
confluence, we have to investigate the properties of this invariant. We introduce
the following definitions from \cite{duckstuckeysulzmann07}. As overlap states
themselves may not satisfy the invariant we have to examine all possible
extensions that satisfy it. Note that in \cite{duckstuckeysulzmann07} CHR states
are defined as 5-tuples consisting of a goal, user store, built-in store, token
store, and the set of global variables. As such a verbose definition is not
necessary for the remainder of this work, we use the more concise state
definition from Section~\ref{sec:prelim:chr} and have adjusted the work from
\cite{duckstuckeysulzmann07} accordingly.

\begin{definition}[Extension, Valid Extension] 

A state~$\sigma = \st{\bbG}{\bbB}{\bbV}$ can be \emph{extended} by another
state~$\sigma_e = \st{\bbG_e}{\bbB_e}{\bbV_e}$ as follows.\[ \sigma \extend
\sigma_e = \st{\bbG \uplus \bbG_e}{\bbB \land \bbB_e}{\bbV_e} \] We say that
$\sigma_e$ is an \emph{extension} of $\sigma$. A \emph{valid
extension}~$\sigma_e$ of a state~$\sigma$ is an extension such that\[ v \in
\vars(\bbG,\bbB) \land v \not \in \bbV \Rightarrow v \not \in \vars(\bbG_e,
\bbB_e, \bbV_e).\]
\end{definition}

When applied to confluence checking with critical pairs there are generally
infinitely many possible extensions of a critical pair. To get a decidable
criterion, the following relation on extensions \footnote{Originally, in
\cite{duckstuckeysulzmann07} this relation is defined as a partial order,
despite being neither transitive nor anti-symmetric. However, it is sufficient
for this work to consider it as a reflexive binary relation.} allows us to
consider only minimal elements.

\begin{definition}[Relation on Extensions]\label{def:partialorder}

Let $\sigma = \st{\bbG}{\bbB}{\bbV}$ be a state, and let $\sigma_{e1} =
\st{\bbG_{e1}}{\bbB_{e1}}{\bbV_{e1}}$ and $\sigma_{e2} =
\st{\bbG_{e2}}{\bbB_{e2}}{\bbV_{e2}}$ be valid extensions of $\sigma$. Then we
define $\sigma_{e1} \preceq_{\sigma} \sigma_{e2}$ to hold if
\begin{enumerate}
  \item there exists a valid extension~$\sigma_{e3}$ of $(\sigma \extend
  \sigma_{e1})$ such that $(\sigma \extend \sigma_{e1}) \extend \sigma_{e3}
  \equiv \sigma \extend \sigma_{e2}$
  \item $\bbV - \bbV_{e1} \subseteq \bbV - \bbV_{e2}$ holds.
\end{enumerate}
\end{definition}

Note that for any extension~$\sigma_e = \st{\bbG_e}{\bbB_e}{\bbV_e}$ of a
state~$\sigma = \st{\bbG}{\bbB}{\bbV}$ there exists a valid extension
$\sigma_{\emptyset} = \st{\emptyset}{\top}{\bbV}$ with $\sigma_{\emptyset}
\preceq_{\sigma} \sigma_e$, simply because the second condition in
Definition~\ref{def:partialorder} is trivially satisfied and $\sigma_{e3} =
\st{\bbG_e}{\bbB_e}{\bbV_e}$ satisfies the first condition.

In the following we want to discuss overlap states that do not satisfy an
invariant~\mcI. Therefore, we are interested in extensions of those states, such
that the result satisfies the invariant~\mcI. The following definition introduces
the set of all those extensions and their minimal elements with respect to the
previously defined relation.

\begin{definition}

Let $\Sigma_e(\sigma)$ be the set of all valid extensions of a state~$\sigma$,
and let $\Sigma^\mcI_e(\sigma) = \{ \sigma_e \mid \sigma_e \in \Sigma_e(\sigma)
\land \mcI(\sigma \extend \sigma_e)\}$ be the set of all valid extensions
satisfying the invariant~\mcI. Finally, let $\mathcal{M}^\mcI_e(\sigma)$ be the
$\prec_{\sigma}$-minimal elements of $\Sigma^\mcI_e(\sigma)$.
\end{definition}

As shown in \cite{duckstuckeysulzmann07} the analysis of critical pairs can be
extended to this context. Instead of requiring joinability of a critical pair --
which might not satisfy the invariant \mcG\ -- we require joinability for all
possible extensions of a critical pair that satisfy \mcG. We make use of the
relation on extensions here, such that we only have to investigate minimal
extensions. Note that we implicitly consider minimal elements modulo built-in
equivalence, e.g., the built-in store~$D=1$ subsumes equivalent stores, like
$D=D'+1 \land D'=0$.

\begin{definition}

A program~\mcP\ is \emph{minimal extension joinable} if for all critical pairs
$\mcCP = (\sigma_1, \sigma_2)$ with overlap~\sigcp, and for all $\sigma_e \in
\mathcal{M}^\mcI_e(\sigcp)$, we have that $(\sigma_1 \extend \sigma_e, \sigma_2
\extend \sigma_e)$ is joinable.
\end{definition}

It has been shown in \cite{duckstuckeysulzmann07} that joinability of critical
pairs, stemming from overlaps with minimal extensions, is a necessary and
sufficient criterion for \mcI-local-confluence if the relation on extensions is
well-founded.

\begin{lemma}[Deciding \mcI-Local-Confluence]\label{lem:obs_confl}

Given that $\prec_{\sigcp}$ is well-founded for all overlaps~$\sigcp$, then:
\mcP\ is \mcI-local-confluent if and only if \mcP\ is minimal extension joinable.
\end{lemma}

Although, in our programs built-in constraints~${+}$ and ${-}$ occur, we can
consider $\prec_{\sigcp}$ well-founded for the following reason: On state
components other than the built-in store the $\prec_{\sigcp}$-relation
corresponds to the well-founded subset ordering with the minimal element
$\emptyset$ (cf. \cite{duckstuckeysulzmann07}). For the built-ins, we can
consider ${+}$ and ${-}$ as successor/predecessor terms (as they are only used
with constants in rules), and hence, we get well-foundedness via proposition~1 of
\cite{duckstuckeysulzmann07}. 

We further note, that for any extension~$\sigma_e$ and state~$\sigcp$ holds that
$\sigma_\emptyset \prec_{\sigcp} \sigma_e$. The following discussion shows that
either $\mcp = \{\sigma_{\emptyset}\}$ or $\Sigma_e^{\mcG}(\sigcp) = \mcp =
\emptyset$. Whether the minimal element $\sigma_{\emptyset}$ exists depends
solely on $\mcG(\sigcp)$ holding as the following lemma shows.

\begin{lemma}[No Minimal Elements]\label{lem:obs_confl_empty_M}

If $\mcG(\sigcp)$ is violated for an overlap~\sigcp\ then no extension~$\sigma_e$
exists such that $\mcG(\sigcp \extend \sigma_e)$ is satisfied, i.e.
$\Sigma_e^{\mcG}(\sigcp) = \mcp = \emptyset$.

\begin{proof}
We proof this by a structural analysis of the overlap which gives the different
possibilities for $\mathcal{G}(\sigma_{\mathcal{CP}})$ to be violated. W.l.o.g.
the overlap stems from the two rules~$\chrrule(r_1) = (r_1\ @\ C_{L_1}
\Leftrightarrow C_{R_1}^u,C_{R_1}^b)$ and $\chrrule(r_2) = (r_2\ @\ C_{L_2}
\Leftrightarrow C_{R_2}^u,C_{R_2}^b)$ with the corresponding rule graphs
$L_1,L_2,K_1,K_2,R_1$, and $R_2$.

First consider the case of nodes~$v_1$ and $v_2$ being overlapped:\\
Let $\type_{L_1}(v_1)(\var(v_1), D_1) \in C_{L_1}$ and
$\type_{L_2}(v_2)(\var(v_2), D_2) \in C_{L_2}$ be overlapped with
$\type_{L_1}(v_1) = \type_{L_2}(v_2)$. The equality constraint $\var(v_1) =
\var(v_2) \in \sigcp$ resembles the merging of the two graph nodes $v_1$ and
$v_2$. However, for the degree equalities different possibilities exist:

\begin{itemize}

  \item $D_1$ and $D_2$ are constants: Then $D_1 = D_2 = \deg_{L_1}(v_1) =
  \deg_{L_2}(v_2) = k$, as the overlap is impossible otherwise. Then \sigcp\
  contains only one constraint~$\type_{L_1}(v_1)(\var(v_1), \deg_{L_1}(v_1))$. As
  in $L_1$ and $L_2$ the nodes each have $k$ adjacent edges, all constraints
  corresponding to adjacent edges in both rule graphs have to be contained in the
  overlap as well. If at least one such constraint is not part of the overlap
  then \sigcp\ contains more than $k$ constraints corresponding to edges adjacent
  to $v_1 = v_2$. As the degree for the node is a constant it cannot be changed
  by any extension and the additional edge constraints cannot be removed either.
  Therefore in such a case, no extension~$\sigma_e$ can correct the degree
  inconsistency and $\mcG(\sigcp \extend \sigma_e)$ cannot hold.

  \item $D_1$ and $D_2$ are variable: In this case the overlap is possible
  without any problems. Depending on the number of overlapped adjacent edge
  constraints the degree variables can always be instantiated with the correct
  degree, thus satisfying the invariant~\mcG.

  \item w.l.o.g. $D_1 = k$ and $D_2$ is a variable: this means $D_2 = k \in
  \sigcp$, therefore, all edge constraints of $C_{L_2}$ of edges adjacent to
  $v_2$ have to be overlapped with edge constraints of $C_{L_1}$ corresponding to
  edges adjacent to $v_1$. If there is such an edge constraint from $C_{L_2}$
  which is not contained in the overlap, then \sigcp\ contains more than $k$ edge
  constraints corresponding to edges adjacent to $v_1$. Again the degree of $v_1$
  is specified as the constant~$k$ in \sigcp, and thus, an extension cannot
  correct this degree inconsistency. If however, all these edge constraints are
  contained in the overlap, \mcG\ is satisfied again, as there are exactly $k$
  such edge constraints coming from $C_{L_1}$.
\end{itemize}

Finally, consider an edge being overlapped:\\
Let $\type_{L_1}(\var(e_1), \var(\src(e_1)), \var(\tgt(e_1))) \in C_{L_1}$ and\\
$\type_{L_2}(\var(e_2), \var(\src(e_2)), \var(\tgt(e_2))) \in C_{L_2}$, then\\
$\var(e_1) = \var(e_2) \land \var(\src(e_1)) = \var(\src(e_2)) \land
\var(\tgt(e_1)) = \var(\tgt(e_2)) \in \sigcp$. By Def.~\ref{def:chr_rule} we have
constraints $\type_{L_1}(\src(e_1))(\var(\src(e_1)), \_) \in C_{L_1}$ and
$\type_{L_2}(\src(e_2))(\var(\src(e_2)), \_) \in C_{L_2}$. If these two
constraints are not part of the overlap, the corresponding equality
constraint~$\var(\src(e_1))=\var(\src(e_2)) \in \sigcp$ results in a single graph
node being represented by two constraints. This is a violation of \mcG, as
$\chr(\gnd, G)$ contains exactly one constraint for each node. This violation
cannot be fixed by an extension, as the conflicting additional node constraint
cannot be removed. Analogously, the two node constraints corresponding to
$\tgt(e_1)$ and $\tgt(e_2)$ have to be contained in the overlap.

Therefore, an overlap~\sigcp\ which violates the invariant~\mcG\ has to violate
it due to one of the above reasons for which it cannot be extended by an
extension~$\sigma_e$ such that $\mcG(\sigcp \extend \sigma_e)$ is satisfied.
\end{proof}

\end{lemma}

Combining these two results yields the criterion in Corollary~\ref{cor:g_confl}
for deciding \mcG-local-confluence. Note that this decision criterion is
essentially the same criterion as used for traditional local confluence, except
that the invariant~\mcG\ restricts the set of investigated overlaps.

\begin{corollary}[Deciding \mcG-Local-Confluence]\label{cor:g_confl}

\mcP\ is \mcG-local-confluent if and only if for all critical pairs~$\mcCP =
(\sigma_1,\sigma_2)$ with overlap~\sigcp, for which $\mcG(\sigcp)$ holds, \mcCP\
is joinable.
\begin{proof}

This follows from the combination of Lemma~\ref{lem:obs_confl},
Lemma~\ref{lem:obs_confl_empty_M} and the insight that $\sigma_{\emptyset}$ is
the unique minimal extension in the case of $\mcG(\sigcp)$ holding.
\end{proof}
\end{corollary}

Next we transfer the joinability of critical CHR pairs to strong joinability in
GTS:

\begin{lemma}[$\mathcal{G}$-Confluence Implies Strong Joinability]
\label{lem:confluence_chr_gts}
If a terminating GTS-CHR program is \mcG-confluent, then all critical GTS pairs
are strongly joinable.

\begin{proof}
Let $P_1 \stackrel{r_1,m_1}{\Longleftarrow} G \stackrel{r_2,
m_2}{\Longrightarrow} P_2$ be a critical GTS pair. Let $r_i = (L_i \leftarrow K_i
\rightarrow R_i)$ and $\chrrule(r_i) = (r_i\ @\ C_{L_i} \Leftrightarrow
C_{R_i}^u,C_{R_i}^b)$ for $i=1,2$.

By Lemma~\ref{lem:cp_gts_chr} there exists an overlap~\sigcp\ which is a
\mcG-state~based~on~$G$. As the critical pair $(\sigma_1,\sigma_2)$ created by
the overlap~\sigcp\ is joinable we have the computations $\sigcp \der \sigma_1
\der^* \tau_1$ and $\sigcp \der \sigma_2 \der^* \tau_2$ with $\tau_1 \equiv
\tau_2$. From Thm.~\ref{thm:sound_and_complete} we know that there exist
corresponding GTS transformations $G \stackrel{r_1,m_1}{\Longrightarrow} P_1
\Longrightarrow^* X_1 \simeq X_2 {}^*\!\!\Longleftarrow P_2
\stackrel{r_2,m_2}{\Longleftarrow} G$. The isomorphism between $X_1$ and $X_2$
follows from Lemma~\ref{lem:eq_iso}. Hence, the critical GTS pair is joinable.

To see that it is strongly joinable consider the set $\strong(\sigcp)$. Every
node~$v$ for which $\track_{G \dergts P_1}(v)$ and $\track_{G \dergts P_2}(v)$
are defined is a node which is not deleted by either $r_1$ or $r_2$. As $m_1$ and
$m_2$ are jointly surjective w.l.o.g. there exists a node~$v' \in V_{L_1}$ of
rule~$r_1$ with $m(v') = v$. As the node is not removed we know $v' \in V_{K_1}$,
and therefore, $\type_{K_1}(v')(\var(v'), \dvar(v')) \in C_{L_1}$. Either the
node is not part of the overlap in $\sigcp$, or if it is overlapped with a
node~$v'' \in V_{L_2}$ such that $m(v') = m(v'')$, then we also know that $v''
\in V_{K_2}$ due to the defined track morphism. Therefore, we always have the
node constraint~$\type_{K_1}(v')(\var(v), \dvar(v)) \in \sigcp$ and $v \in
\strong(\sigcp)$. As this node cannot be removed during the transformation, a
variant of this constraint with adjusted degree is also present in $\tau_1$ and
$\tau_2$. These two variant constraints are uniquely determined, as $\var(v) \in
\bbV$ by Def.~\ref{def:chr_cp}, and hence, they both have to use $\var(v)$ for
the node identifier variable. This means we still have to show for such a
node~$v$ that the two conditions from Def.~\ref{def:gts_cp} are satisfied:
\begin{enumerate}
  \item $\track_{G \dergts P_1 \dergts X_1}(v)$ and $\track_{G
  \dergts P_2 \dergts X_2}(v)$ are defined:\\
  By Thm.~\ref{thm:sound_and_complete} we know that the GTS transformations are strong
  w.r.t. $\strong(\sigcp)$. As $v \in \strong(\sigcp)$ this implies $v \in m(K)
  \lor v \not \in m(L)$ for each of the applied rules, i.e. the node remains
  during the transformation and hence the track morphisms are defined as in
  Def.~\ref{def:track}.

  \item $f_V(\track_{G \dergts P_1 \dergts X_1}(v)) = \track_{G
  \dergts P_2 \dergts X_2}(v)$:\\
  As the isomorphism~$f$ is derived from $\tau_1 \equiv \tau_2$ and $\var(v) \in
  \bbV$ this isomorphism correctly relates the original node~$v$ with its
  occurrences in $\tau_1$, resp. $X_1$, and $\tau_2$, resp.
  $X_2$.\end{enumerate}\end{proof}
\end{lemma}

The reverse direction holds as well, as the following lemma shows.

\begin{lemma}[Strong Joinability Implies
\mcG-Confluence]\label{lem:confluence_gts_chr} 

If all critical GTS pairs of a terminating GTS are strongly joinable, then the
corresponding GTS-CHR program is \mcG-confluent.
\begin{proof}
Consider an overlap~\sigcp\ for the critical CHR pair~$(\sigma_1,\sigma_2)$.
W.l.o.g. $\mcG(\sigcp)$ holds according to Cor.~\ref{cor:g_confl}.
Therefore, $\sigcp$ is a \mcG-state~based~on~$G$ and $\sigma_1,
\sigma_2$ correspond to graphs~$G_1,G_2$. Consider now $G_1
\stackrel{r_1,m_1}{\Longleftarrow} G \stackrel{r_2,m_2}{\Longrightarrow} G_2$.

We now show, that either the critical CHR pair is non-critically joinable, or it
corresponds to a critical GTS pair and can thus be joined, because all critical
GTS pairs are strongly joinable.

First, we want to point out that $G$ is minimal by the definition of the CHR
overlap, i.e. every occurring node and edge is part of a match, hence, $m_1$ and
$m_2$ are jointly surjective.

Next, we distinguish two cases: First, let $G_1
\stackrel{r_1,m_1}{\Longleftarrow} G \stackrel{r_2,m_2}{\Longrightarrow} G_2$ be
parallel independent. Therefore, the second rule can be applied after the
first, because none of the required nodes or edges has been removed. The
following diagram depicts this situation:\\
\centerline{
\xymatrix{
& \ar@{=>}[dl]_{r_1} G \ar@{=>}[dr]^{r_2} & \\
G_1 \ar@{=>}[dr]^{r_2} & & \ar@{=>}[dl]_{r_1} G_2 \\
& X &
}}

By Theorem~\ref{thm:sound_and_complete} we can apply the corresponding rules to
$\sigcp$ in order to join the critical CHR pair, because $\strong(\sigcp)$
contains only nodes not deleted by $r_1$ and $r_2$.

Secondly, let $G_1 \stackrel{r_1,m_1}{\Longleftarrow} G
\stackrel{r_2,m_2}{\Longrightarrow} G_2$ be parallel dependent. It follows that
$m(L_1) \cap m(L_2) \not \subseteq m(K_1) \cap m(K_2)$. However, this is now a
critical GTS pair, and hence, strongly joinable as depicted on the left of the
following diagram:\\
\centerline{
\xymatrix{
& \ar@{=>}[dl]_{r_1} G \ar@{=>}[dr]^{r_2} & & & \ar[dl]_{r_1} \sigcp
\ar[dr]^{r_2}&\\
G_1 \ar@{=>}[dr]^{*} & (GTS) & \ar@{=>}[dl]_{*} G_2 &  \sigma_1\ar[dr]^{*}
&(CHR)& \ar[dl]_{*}\sigma_2 \\
& X_1 \simeq X_2 & & & \sigma_1' \equiv \sigma_2' &
}}

The right part of the diagram shows the situation for the critical CHR pair which
is joinable by Thm.~\ref{thm:sound_and_complete}. This is possible, because
$\forall v \in \strong(\sigcp)$ we know that $\track_{G\dergts G_1}(v)$ and
$\track_{G \dergts G_2}(v)$ are defined, thus by Def.~\ref{def:gts_cp}, $v$
is never removed and still present in $X_1$ and $X_2$. Finally, the isomorphism
implied by $X_1 \simeq X_2$ gives us $\sigma_1' \equiv \sigma_2'$. Note that
despite Lemma~\ref{lem:eq_iso} not being reversible in general this holds here,
as it is clearly determined for both $\sigma_1'$ and $\sigma_2'$ which node
identifier variables are global and the strong joinability condition reflects
this in the isomorphism.

Therefore, for all overlaps~$\sigcp$ with $\mcG(\sigcp)$ holding we know that the
corresponding critical CHR pair is joinable, and hence, by Cor.~\ref{cor:g_confl}
that the CHR program is $\mathcal{G}$-local-confluent. As it is terminating as
well, it is $\mathcal{G}$-confluent.
\end{proof}
\end{lemma}

The combination of the previous two lemmata gives us our main result:

\begin{theorem}[Strong Joinability iff
\mcG-Confluence]\label{thm:confluence_characterization}

All critical GTS pairs of a terminating GTS are strongly joinable if and only if
the corresponding GTS-CHR program is \mcG-confluent.
\begin{proof}
Direct combination of Lemma~\ref{lem:confluence_chr_gts} and
Lemma~\ref{lem:confluence_gts_chr}.
\end{proof}
\end{theorem}

\begin{corollary}[\mcG-Confluence Implies GTS Confluence]

If a terminating GTS-CHR program is \mcG-confluent, then the corresponding GTS is
confluent.
\begin{proof}
Strong joinability is a sufficient criterion for confluence of a terminating
GTS (cf. \cite{plump05}). Therefore, this follows directly from
Theorem~\ref{thm:confluence_characterization}.
\end{proof}
\end{corollary}

Practically, with Theorem~\ref{thm:confluence_characterization} we can reuse the
automatic confluence check for terminating CHR programs
\cite{abdennadherfruehwirthmeuss99,fruehwirth09} to prove confluence of a
terminating GTS-CHR program. As Lemma~\ref{lem:obs_confl_empty_M} showed, it is
sufficient to only consider overlaps satisfying the graph invariant~\mcG.
Whenever all the resulting critical CHR pairs are joinable, the CHR program is
\mcG-confluent according to Corollary~\ref{cor:g_confl}. This, in turn, is
sufficient for proving confluence of the original GTS.

\subsection{Discussion}
\label{sec:confl_discussion}

In this section we elaborate on some canonical examples that highlight different
properties of critical pairs. These examples are inspired by \cite{plump05}.

\begin{example}

Consider the following rules which use two different edge types: a and b
\begin{center} 
\includegraphics{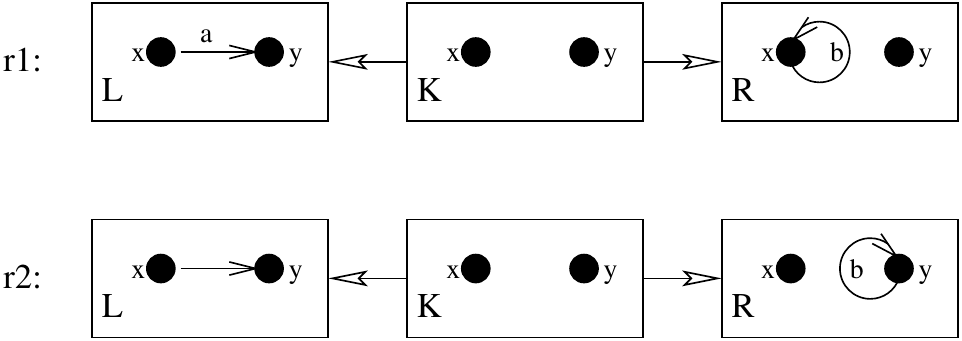}
\end{center}

The only critical GTS pair of these rules is joinable. This is possible in the
GTS case, because the resulting graphs, shown below, are isomorphic.

\begin{center}
\includegraphics{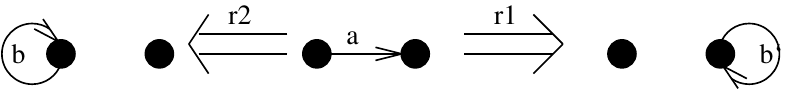} 
\end{center}

However, the track morphisms of the above derivations are incompatible, i.e. the
strong joinability condition from Definition~\ref{def:gts_cp} cannot be
satisfied. As the following derivation shows, this hinders monotonicity and
joinability is lost, when the critical pair is embedded into a larger context.

\begin{center}
\includegraphics{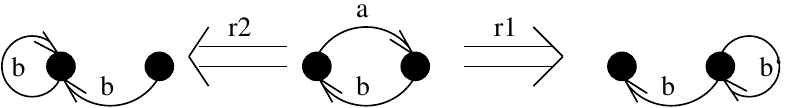} 
\end{center}

The two resulting states are no longer isomorphic and also cannot be joined, as
no more rules are applicable to them. Therefore, this GTS is not locally
confluent, although all its critical GTS pairs are joinable.

We now examine this scenario in CHR. The two GTS rules then become the
following CHR rules:

\begin{center} 
\begin{tabular}{ll}
$r1$ @ & $\node(N_x, D_x) \uplus \node(N_y, D_y) \uplus \text{a}(E, N_x, N_y)$
\\ & $\Leftrightarrow$ \\
& $\node(N_x,D_x') \uplus \node(N_y, D_y') \uplus \text{b}(E', N_x, N_x), $\\
& $D_x' = D_x{+}1 \land D_y' = D_y{-}1$\\
$r2$ @ & $\node(N_x, D_x) \uplus \node(N_y, D_y) \uplus \text{a}(E, N_x,
N_y)$\\ & $\Leftrightarrow$ \\
& $\node(N_x,D_x') \uplus \node(N_y, D_y') \uplus \text{b}(E', N_y, N_y),$\\
& $D_x' = D_x{-}1 \land D_y' = D_y{+}1$
\end{tabular}
\end{center}

We now consider the critical CHR pair corresponding to the above critical GTS
pair. It is generated by fully overlapping both rule heads, resulting in the
overlap \[\sigcp = \st{\node(N_1, D_1)\uplus\node(N_2, D_2)\uplus\text{a}(E,
N_1, N_2)}{\top}{\bbV}\] with $\bbV = \{N_1, N_2, D_1, D_2, E\}$. The resulting
critical CHR pair~$(\sigma_1, \sigma_2)$ is:
\begin{eqnarray*} 
\st{\node(N_1,D_1') \uplus \node(N_2, D_2') \uplus b(E', N_1,
N_1)}{D_1' = D_1{+}1 \land D_2' = D_2{-}1}{\bbV},\\
\st{\node(N_1,\tilde{D_1}) \uplus \node(N_2, \tilde{D_2}) \uplus
b(\tilde{E}, N_2, N_2)}{\tilde{D_1} = D_1{-}1 \land \tilde{D_2} =
D_2{+}1}{\bbV}
\end{eqnarray*}
It is clear that $\sigma_1 \not \equiv \sigma_2$, because $\CT \not \models (D_1' =
D_1{+}1 \land D_2' = D_2{-}1) \rightarrow \exists_\emptyset N_1 = N_2$ as
required by Theorem~\ref{thm:equiv_tf}.

The strong nodes~$N_1$ and $N_2$, i.e. $N_1,N_2 \in \bbV$, enforce compatible
track morphisms, and hence are responsible for the non-joinability above. If we
instead want to test non-strong joinability, we can do so as well by setting
$\bbV = \emptyset$. Then, the two states $\sigma_1$ and $\sigma_2$ are indeed
equivalent by Definition~\ref{def:equiv}, as $N_2$ is existentially quantified
and the remaining conditions of Theorem~\ref{thm:equiv_tf} hold as well.
\end{example}

\begin{example}\label{ex:counterexample}
Another example from \cite{plump05} is the following GTS which is terminating and
confluent, however, the critical GTS pair from the overlap of rule~$r1$ with
itself is not strongly joinable. This is a counterexample used to show that
strong joinability of critical GTS pairs is only a sufficient criterion for
confluence of a terminating GTS.

\begin{center} 
\includegraphics{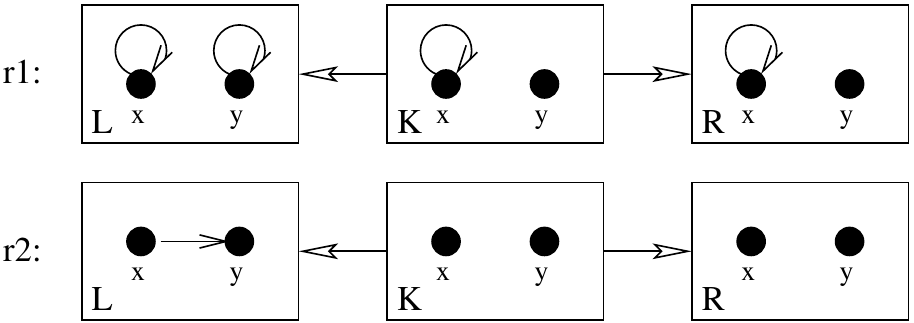}
\end{center}

The GTS works as follows: If there is at least one loop in the graph, then all
but a last loop are removed by the first rule. Additionally, all non-loop edges
are removed by the second rule. Therefore, the remaining final graph contains
zero or one loops and no other edges, and hence the GTS is terminating and
confluent due to graph isomorphism. The first rule is encoded in CHR as follows:

\begin{center} 
\begin{tabular}{ll}
$r1$ @ & $\node(N_x, D_x) \uplus \node(N_y, D_y) \uplus$\\
& $\text{a}(E_x, N_x, N_x) \uplus \text{a}(E_y, N_y, N_y)$ \\
& $\Leftrightarrow$ \\
& $\node(N_x, D_x) \uplus \node(N_y, D_y') \uplus \text{a}(E_x, N_x, N_x),$\\
& $D_y' = D_y - 2$
\end{tabular}
\end{center}

Completely overlapping the rule with itself yields the overlap \[\sigcp =
\st{\node(N_1, D_1) \uplus \node(N_2, D_2) \uplus \text{a}(E_1, N_1, N_1) \uplus
\text{a}(E_2, N_2, N_2)}{\top}{\bbV}\] with $\bbV = \{N_1, N_2, D_1,
D_2, E_1, E_2\}$ resulting in the critical CHR pair~$(\sigma_1, \sigma_2)$ with:
\[
\begin{array}{l}
\sigma_1 = \st{\node(N_1, D_1) \uplus \node(N_2, D_2') \uplus \text{a}(E_1, N_1,
N_1)}{D_2' = D_2 - 2}{\bbV}\\
\sigma_2 = \st{\node(N_1, D_1') \uplus \node(N_2, D_2)\uplus \text{a}(E_2, N_2,
N_2)}{D_1' = D_1 - 2}{\bbV}
\end{array}
\]

Analogously to the previous example, the two states are not equivalent and cannot
be joined, therefore the corresponding critical GTS pair is not strongly
joinable. Again, setting $\bbV = \emptyset$ results in both states becoming
equivalent. As before, this reflects that for the critical GTS pairs the two
corresponding graphs are isomorphic.
\end{example}

\section{Analyzing Operational Equivalence}
\label{sec:opeq}

Constraint Handling Rules is well-known for its decidable, sufficient, and
necessary criterion for operational equivalence of terminating and confluent
programs \cite{abdennadherfruehwirth99,fruehwirth09}. After presenting this
result in Section~\ref{sec:opeq_chr}, we introduce the concept of operational
equivalence for graph transformation systems in Section~\ref{sec:opeq_gts}. Then
we investigate operational equivalence of GTS-CHR programs and show that it is
sufficient for operational equivalence of the original GTS. We further
demonstrate its application to detect redundant rules of a GTS.

The contents of this section are a revised and extended version of
\cite{Raiser2009b}.

\subsection{Operational Equivalence in CHR}
\label{sec:opeq_chr}

Operational equivalence, intuitively, means that two programs should be able to
compute equivalent outputs given the same input. Applied to a single state, this
behavior is called \Ps-joinability:

\begin{definition}[\Ps-joinability]\label{def:p_joinable} Let $\Ps$ be CHR
programs. A state~$\sigma$ is \emph{\Ps-joinable}, if and only if there are
computations $\sigma \der_{\mcP_1}^* \sigma_1$ and $\sigma \der_{\mcP_2}^*
\sigma_2$ with $\sigma_1 \equiv \sigma_2$ where all $\sigma_i$ are final states
with respect to $\mcP_i$.
\end{definition}

If \Ps-joinability is given for all states the programs are considered
operationally equivalent:

\begin{definition}[Operational Equivalence]
Let $\Ps$ be CHR programs.

$\Ps$ are \emph{operationally equivalent} if and only if all states~$\sigma$
are \Ps-joinable.
\end{definition}

As mentioned before, operational equivalence is decidable for terminating and
confluent CHR programs. Similarly to confluence, the decision algorithm
investigates critical states created from rule heads.

\begin{definition}[Critical States]\label{def:critical_states} 
Let $\Ps$ be CHR programs. The set of \emph{critical states of $\mcP_1$ and
$\mcP_2$} is defined as $\{ \st{H}{\top}{\vars(H)} \mid (H \Leftrightarrow
B_c,B_b) \in \mcP_1 \cup \mcP_2 \}$.
\end{definition}

Note that we had to consider observable confluence for CHR, because overlap
states constructed for critical pair analysis may not always encode a graph. The
critical states used for operational equivalence here, however, stem directly
from a complete head of a rule, which in turn was derived from a GTS rule graph.
Therefore, all critical states of GTS-CHR programs are valid encodings of
graphs.

The following theorem, adapted from \cite{abdennadherfruehwirth99}, is based on
the idea to determine \Ps-joinability of these critical states. The monotonicity
property of CHR ensures, that if all critical states are \Ps-joinable, then all
states are. Additionally demanding termination and confluence of the programs,
allows us to decide \Ps-joinability simply by executing a critical state in each
of the programs and then comparing the resulting final states.

\begin{theorem}[Operational Equivalence via Critical States]\label{thm:opeq} Let
$\Ps$ be terminating and confluent CHR programs. $\Ps$ are operationally
equivalent if and only if for all critical states~$\sigma$ of $\mcP_1$ and
$\mcP_2$ it holds that $\sigma$ is \Ps-joinable.
\begin{proof}
Given in \cite{abdennadherfruehwirth99}.
\end{proof}
\end{theorem}

Note that in contrast to confluence, Theorem~\ref{thm:opeq} will always consider
states satisfying the \mcG-invariant, when applied to a GTS-CHR program. This
follows from the fact, that each critical state is the head of a rule, and in
turn, corresponds to a rule graph from the GTS by construction.

\subsection{Analyzing Operational Equivalence in GTS}
\label{sec:opeq_gts}

In this section we introduce the notion of operational equivalence for GTS. Based
on the previous embedding of GTS in CHR, we use the existing decision algorithm
from CHR as a sufficient criterion for operational equivalence of two graph
transformation systems.

First, we define the property of \Ss-joinability for two graph transformation
systems \Ss, analogously to \Ps-joinability.

\begin{definition}[\Ss-joinability]

Let \Ss\ be two graph transformation systems. A typed graph~$G$ is
\emph{\Ss-joinable} if and only if there are derivations $G \dergts^*_{\mcS_1}
G_1$ and $G \dergts^*_{\mcS_2} G_2$ with $G_1 \iso G_2$ being final with respect
to $\mcS_1$ and $\mcS_2$. 

Here $\iso$ denotes traditional graph isomorphism and a graph~$G$ is considered
\emph{final} with respect to $\mcS$ iff there is no transition~$G \dergts_{\mcS}
H$ for any graph~$H$.
\end{definition}

Building on \Ss-joinability, we now define operational equivalence for graph
transformation systems with the same intuitive understanding: two operationally
equivalent GTS should be able to produce the same result graphs up to isomorphism
given an input graph:

\begin{definition}[GTS Operational Equivalence]\label{def:gts_oe}
Let $\mcS_1 = (\mcP_1, TG)$ and $\mcS_2 = (\mcP_2, TG)$ be two graph
transformation systems.

\Ss\ are \emph{operationally equivalent} if and only if for all graphs~$G$ typed
over $TG$ it holds that $G$ is \Ss-joinable.
\end{definition}

Similar to operational equivalence in CHR, where it is futile to directly compare
programs that use different constraints, Definition~\ref{def:gts_oe} requires
$\mcS_1$ and $\mcS_2$ to be based on the same type graph~$TG$. With the previous
results from \cite{Raiser2009} we can directly use CHR's operational equivalence
as a sufficient criterion for deciding operational equivalence of two GTS:

\begin{theorem}[GTS-CHR Operational Equivalence]\label{thm:gtsopeq}
Let \Ss\ be graph transformation systems and \Ps\ their corresponding GTS-CHR
programs. \Ss\ are operationally equivalent if \Ps\ are operationally
equivalent.

\begin{proof}

Let $G$ be a graph typed over $TG$. Then the state~$\sigma = \st{\chr(\gnd,
G)}{\top}{\emptyset}$ is \Ps-joinable by Def.~\ref{def:p_joinable}. Therefore,
there exist the final states~$\sigma_1 \equiv \sigma_2$ with $\sigma
\der_{\mcP_1}^* \sigma_1$ and $\sigma \der_{\mcP_2}^* \sigma_2$.

By Thm.~\ref{thm:sound_and_complete} we know that there exist corresponding
derivations~$G \dergts_{\mcS_1}^* G_1$ and $G \dergts_{\mcS_2}^* G_2$ such
that $\sigma_1$ is a \mcG-state~based~on~$G_1$ and $\sigma_2$ is a
\mcG-state~based~on~$G_2$.

The graphs~$G_1$ and $G_2$ are final states w.r.t. $\mcS_1$ and $\mcS_2$, and
finally, the isomorphism between $G_1$ and $G_2$ is implied by $\sigma_1 \equiv
\sigma_2$ according to Lemma~\ref{lem:eq_iso}. Therefore, $G$ is \Ss-joinable.
\end{proof}
\end{theorem}

An interesting application of the above theorem is the removal of redundant
rules. Originally proposed in \cite{Abdennadher2003}, decidable operational
equivalence of CHR programs implies a straight-forward redundant rule removal
algorithm: Remove a single rule from the program, then compare the operational
equivalence of the program thus created and the original program. If the two
programs are operationally equivalent the selected rule is shown to be redundant
and can be removed.

Clearly, program equivalence in general is undecidable, and hence, we cannot
expect such an algorithm to correctly identify all redundant rules. Nevertheless,
the algorithm was applied in CHR research to great success on automatically
generated programs \cite{Abdennadher2007,raiser08cp}. These generations tend to
create rules which subsume each other, in which case the algorithm works well as
the following example demonstrates.

\begin{figure}
\centerline{
\scalebox{.8}{\includegraphics{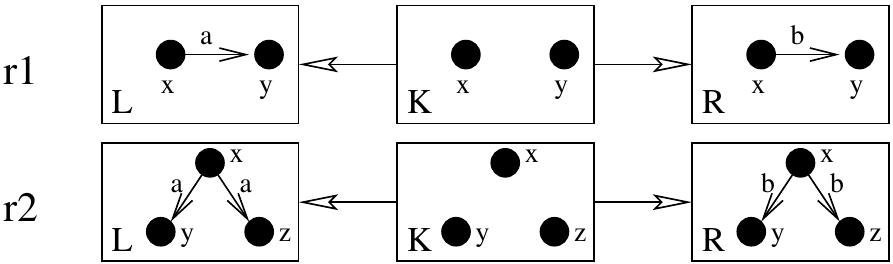}}
}
\caption{Example of a graph transformation system}
\label{fig:ex:opeq}
\end{figure}

\begin{example}

Consider the graph transformation system~$\mcS_1$ given in
Figure~\ref{fig:ex:opeq}. It depicts a typical case, in that the rule~$r2$ is
subsumed by rule~$r1$. While this is easily verified by a human reader,
Theorem~\ref{thm:opeq} gives us the means for an automated verification.

In order to verify the redundancy of rule~$r2$, consider a second graph
transformation system~$\mcS_2$, which contains only rule~$r1$. Proving that
$\mcS_1$ and $\mcS_2$ are operationally equivalent then proves the redundancy of
rule~$r2$.

Encoding the graph transformation system~$\mcS_1$ from Figure~\ref{fig:ex:opeq}
in CHR results in the following two rules:\\
\begin{tabular}{ll}
$r1$ @ & $\node(N_x, D_x) \uplus \node(N_y, D_y) \uplus \text{a}(E, N_x, N_y)$
\\
& $\Leftrightarrow$ \\
& $\node(N_x, D_x) \uplus \node(N_y, D_y) \uplus \text{b}(E', N_x, N_y)$\\
\\
$r2$ @ & $\node(N_x, D_x) \uplus \node(N_y, D_y) \uplus \node(N_z, D_z)
\uplus$\\
& $\text{a}(E_y, N_x, N_y) \uplus \text{a}(E_z, N_x, N_z)$ \\
& $\Leftrightarrow$ \\
& $\node(N_x, D_x) \uplus \node(N_y, D_y) \uplus \node(N_z, D_z) \uplus$\\
& $\text{b}(E_1, N_x, N_y) \uplus \text{b}(E_2, N_x, N_z)$
\end{tabular}

This GTS-CHR program~$\mcP_1$ is confluent and terminating and the same holds for
$\mcP_2$, which encodes $\mcS_2$ respectively. Next, we investigate the
$\Ps$-joinability of all critical states, of which there are two. The critical
state derived from $r1$ is clearly $\Ps$-joinable, as the same rule can be
applied to it in both programs, resulting in equivalent final states.

The critical state derived from rule~$r2$ contains two a-edges, which can be
converted to b-edges either by applying rule~$r2$ or rule~$r1$ twice. Therefore,
the final states in both programs are equivalent again, and hence, the programs
are operationally equivalent. As a conclusion, $\mcS_1$ and $\mcS_2$ are
operationally equivalent, which in turn proves the redundancy of rule~$r2$.

\end{example}

In general, Theorem~\ref{thm:gtsopeq} cannot be reversed, i.e. it is only a
sufficient, not a necessary criterion. A counterexample for the reverse direction
is given in the following example. Notice that it is based on the example used by
\cite{plump05} in order to demonstrate why the critical pair lemma is not a
necessary criterion for confluence. This might be seen as an indication that a
similar situation exists for GTS program equivalence.

\begin{example}

Consider two GTS with the first being the one from
Example~\ref{ex:counterexample} and the second GTS is identical to the first
except for rule~$r1$, in which the loop for node~$x$ is removed instead. It is
clear, that both programs are terminating, confluent, and operationally
equivalent. The following two rules are from the corresponding GTS-CHR
programs~$\mcP_1$ and $\mcP_2$:\\
\begin{tabular}{ll}
$r1$ @ & $\node(N_x, D_x) \uplus \node(N_y, D_y) \uplus$\\
& $\text{a}(E_x, N_x, N_x) \uplus \text{a}(E_y, N_y, N_y)$ \\
& $\Leftrightarrow$ \\
& $\node(N_x, D_x) \uplus \node(N_y, D_y') \uplus \text{a}(E_x, N_x, N_x),$\\
& $D_y' = D_y {-} 2$\\
\\
$r1'$ @ & $\node(N_x, D_x) \uplus \node(N_y, D_y) \uplus$\\
& $\text{a}(E_x, N_x, N_x) \uplus \text{a}(E_y, N_y, N_y)$ \\
& $\Leftrightarrow$ \\
& $\node(N_x, D_x') \uplus \node(N_y, D_y) \uplus \text{a}(E_y, N_y, N_y),$\\
& $D_x' = D_x{-}2$
\end{tabular}

We can now investigate the following critical state~$\sigma$ according to
Theorem~\ref{thm:gtsopeq}, where $\bbV = \{N_x,N_y,D_x,D_y,E_x,E_y\}$:
\[
\sigma = \st{\node(N_x, D_x) \uplus \node(N_y, D_y) \uplus \text{a}(E_x, N_x,
N_x) \uplus \text{a}(E_y, N_y, N_y)}{\top}{\bbV}
\]

The critical state~$\sigma$ is not \Ps-joinable, as there is only one rule
applicable in each program and the resulting states are not equivalent:\\
\begin{tabular}{r} 
$\sigma \der_{\mcP_1}^{r1} \st{\node(N_x, D_x) \uplus \node(N_y, D_y') \uplus
\text{a}(E_x, N_x, N_x)}{D_y' = D_y {-} 2}{\bbV} = \tau_1$\\
$\not \equiv$\\
$\sigma \der_{\mcP_2}^{r1'} \st{\node(N_x, D_x') \uplus \node(N_y, D_y) \uplus
\text{a}(E_y, N_y, N_y)}{D_x' = D_x{-}2}{\bbV} = \tau_2$
\end{tabular}
\end{example}   

\section{Related and Future Work}
\label{sec:related_work}

The relation of CHR to other formalisms has been thoroughly investigated in the
literature. This includes comparison to logical formalisms (e.g., linear logic
\cite{Betz2005}), term rewriting
(\cite{duck_stuck_brand_acd_term_rewriting_iclp06}), Join-Calculus
(\cite{lam_sulz_finallyjoin_chr08}), and Petri nets \cite{betz_petri_nets_chr07}.
More detailled surveys of these relations can be found in
\cite{chr_survey_tplp08} and \cite{fruehwirth09}.

The relation of graph transformation systems to CHR differs from these other
formalisms, because firstly, it is a graph-based formalism, and secondly, there
are significant differences in program analysis results. Most importantly,
confluence of terminating GTS is undecidable \cite{plump05} whereas confluence of
terminating CHR programs is decidable \cite{abdennadherfruehwirthmeuss99}.
Furthermore, no operational equivalence analysis exists for GTS, as opposed to
the situation in CHR \cite{abdennadherfruehwirth99}.

The operational equivalence test presented in Section~\ref{sec:opeq} yields a
method for removal of redundant rules, which is remarkable for another reason: In
\cite{Kreowski2000} the notions of redundancy and subsumptions have been
introduced for GTS, however, the authors only gave the definitions and a
sufficient condition for redundancy, but no verification procedure. While the
notion of redundancy in that paper is slightly different from the one found in
\cite{Abdennadher2003}, the adaptation of the algorithm to GTS-CHR programs is to
the best of our knowledge the only available verification procedure for redundant
GTS rules.

Note, that operational equivalence, as defined here, is only one possible notion
of equivalence between programs. It was used in this work as an example of CHR
program analyses applied to embedded graph transformation systems. Another, more
popular, notion of equivalence of GTS is bisimilarity, introduced in the GTS
context in \cite{Ehrig2004}. It has been successfully applied to determine
behavioral equivalence of graph transformation systems in \cite{Rangel2008}.
While bisimilarity originated from process calculi and is focused on the
transitions made during computations of a result, operational equivalence on the
other hand, only compares the final computational results, independently of how
they are reached.

The encoding of GTS in CHR, as presented in Section~\ref{sec:encoding}, is based
on the double-pushout approach for graph transformation systems. A related graph
rewriting mechanism, the single-pushout approach, was introduced in
\cite{loewe93}. Instead of demanding two pushouts, as in Figure~\ref{fig:dpo},
rewriting is defined there over a category of partial graph morphisms, hence only
a single pushout construction is used. Intuitively, this results in a different
behavior with respect to dangling edges: While the double-pushout approach
prohibits a rule application in case a dangling edge would remain, the
single-pushout approach removes all dangling edges instead. In \cite{Lowe1993}
the authors investigate confluence for single-pushout graph rewriting. In
particular, the critical pair analysis is shown to be only a sufficient criterion
as well, not a necessary one.

In this work, we based our encoding on the DPO approach as the non-applicability
of rules due to the dangling edge condition corresponds nicely to
non-applicability of corresponding CHR rules. In order to support the approach
from \cite{loewe93}, remaining dangling edges would need to be removed by an
additional rule, hence, we would lose the one-on-one correspondence of GTS and
CHR rules.

Our encoding further serves as the foundation of the extensible platform for the
analysis of graph transformation systems using constraint handling rules
presented in the diploma thesis \cite{Wasserthal2009}. This platform is based on
JCHR \cite{jchr}, a Java-based implementation of CHR and the work presented in
Section~\ref{sec:encoding}. The developed tool presents a graphical view of a GTS
which is synchronized with the corresponding GTS-CHR program at all times.
Furthermore, it provides an interface for program analysis plug-ins, which can
work directly on the GTS or on the GTS-CHR program.

As this work demonstrated, our embedding leads to cross-fertilizations of CHR and
GTS research. Future work should therefore concentrate on further comparing the
different approaches to program analysis. In particular, CHR provides several
approaches to termination analysis
\cite{Fruhwirth2000,voets_pilozzi_deschreye_termination_techrep08,Pilozzi2008}
that GTS research may profit from.

Research on GTS contains several extensions for the typed graphs and rules
considered in this work. One such extension adds attributes
\cite{ehrigprangetaentzer06} to graphs, which can then be modified by rules. We
assume that built-in constraints available in CHR could closely correspond to
attributes. Another important extension, is the addition of negative application
conditions \cite{ehrigprangetaentzer06}, i.e. applying a rule requires the
absence of certain graph structures. This is more difficult to achieve in CHR, as
it traditionally has no support for negation as absence. However, there exist
proposed extensions of CHR with negation as absence
\cite{vanweert_sney_schr_demoen_negation_chr06} and aggregates
\cite{vanweert_sney_demoen_aggregates_lopstr07}, which could help in extending
our encoding to allow application conditions.

Our work on operational equivalence for graph transformation systems yielded a
first useable criterion. However, there is lot of remaining work in this field.
From a decidability point of view, operational equivalence is in a similar
situation as confluence: \cite{plump05} showed that confluence is undecidable
even for terminating GTS and we expect a similar result for operational
equivalence. Therefore, our criterion might only be applicable to a small subset
of all GTS.

Similarly, in CHR research, the operational equivalence result assumes that both
programs use the same constraint symbols in the same manner. While this
restriction yields a decidable criterion, it also means that it seldomly applies
to real-world programs. Traditionally, one may be able to manually show
operational equivalence for two concrete programs by taking into account known
restrictions on data structures or inputs and ignoring irrelevant states. The
same situation was present for confluence (e.g., \cite{fruehwirth05}) until
observable confluence \cite{duckstuckeysulzmann07} succeeded in providing an
extended approach.

We plan to develop such an invariant-based approach for operational equivalence
in CHR as well, which extends Theorem~\ref{thm:opeq}. Combined with a better
criterion for operational equivalence in GTS, including the track morphism
similarly to the critical pair approach, this might reveal a closer
correspondence between operational equivalence in both systems.

\section{Conclusion}
\label{sec:conclusion}

We have shown that constraint handling rules (CHR) provides an elegant way for
embedding graph transformation systems (GTS). The resulting rules are concise and
directly related to the corresponding graph production rules. We proved soundness
and completeness of this embedding and verified formal properties of CHR states
that encode graphs. Furthermore, we considered partial graphs and showed that the
CHR embedding naturally supports these, hence facilitating program analysis.

Next, we analyzed confluence and showed that observable confluence of a GTS-CHR
program is a sufficient criterion for confluence of the analyzed GTS.
Furthermore, we transferred the notion of operational equivalence from CHR to GTS
and discussed the CHR-based decision algorithm for redundant rule removal.

\bibliography{bibliography}

\end{document}